\documentclass[a4paper,12pt]{article}

\usepackage{amsmath, amsthm, amsfonts, amssymb}
\usepackage{graphicx}

\def\sh{\sinh}
\def\ch{\cosh}

\theoremstyle{plain}
\newtheorem{theorem}{Theorem}[section]

\newtheorem{lemma}{Lemma}[section]

\theoremstyle{remark}
\newtheorem{remark}{Remark}[section]

\def\e{\varepsilon}

\numberwithin{equation}{section}

\begin{document}
\allowdisplaybreaks


\begin{center}

\textbf{\Large Discrete spectrum of a pair of nonsymmetric
waveguides coupled by a window}

\bigskip

{\large D.I. Borisov}\footnotetext[1]{The work is partially
supported by RFBR and the program of supporting leading
scientific schools. The author is also supported by Marie Curie
International Fellowship of 6th European Community Framework
(MIF1-CT-2005-006254) }

\begin{quote}
\emph{Bashkir State Pedagogical University, October rev. st.,
3a, \\ Ufa, Russia, 450000. E-mail:}
\texttt{borisovdi@yandex.ru}
\\ \emph{URL: }\texttt{http://borisovdi.narod.ru/}
\end{quote}

\end{center}

\begin{abstract}

In the paper we study the discrete spectrum of a pair of quantum
two-dimensional waveguides having common boundary in which a
window of finite length is cut out. We study the phenomenon of
new eigenvalues emerging from the threshold of the essential
spectrum when the length of window passes through critical
values. We construct the asymptotics expansions for the emerging
eigenvalues with respect to small parameter which is the
difference between current length of the window and the nearest
critical value. We also study the behaviour of the spectrum when
the length of the window increases unboundedly and construct
asymptotics expansions with respect to great parameter which is
a length of the window.

\end{abstract}

\section*{Introduction}

In last years much attention was paid to the study of spectral
properties of the elliptic operators in unbounded domains with
various perturbations. First of all this is due various
applications of such problem in quantum mechanics and acoustics.
Moreover, these problems possess various features interesting
from mathematical point of view. One of such examples is a
problem on bound states of two quantum waveguides coupled by a
window. Mathematically this corresponds to an eigenvalue problem
for the Dirichlet Laplacian in a domain formed by two parallel
strips having common boundary in which a window of finite length
is cut out (cf. figure). Such model was suggested in the paper
\cite{ESTV}; physical aspects of this problem were discussed
there as well (see also \cite{HTW}). Besides, in  \cite{ESTV}
the authors obtained two-sided estimates for the eigenvalues and
proved that the presence of the window leads to a non-empty
discrete spectrum, the number of isolated eigenvalues increases
when the length of the window does, eigenvalues appear when the
length of the window passes through some critical values. A
number of numerical results was obtained as well. The existence
of at least one isolated eigenvalue in the case of the same
widths of the strips was proved independently in  \cite{BGRS}.
For a sufficiently small window this system has exactly one
isolated eigenvalue. In the case of symmetric strips a two-sided
estimate was obtained for this eigenvalue in \cite{EV1}. In
\cite{EV2} similar result was established for several windows
and non-symmetric strips as well as for two parallel layers
coupled by a window. The case of small window was also
considered in \cite{P}, where the asymptotics expansion for the
aforementioned eigenvalue was formally constructed. The rigorous
proof of the asymptotics expansions in the case of small window
was adduced recently in \cite{G}. In \cite{BEG} the case of the
strips of the same width and finite window was treated. The
phenomenon of new eigenvalues emerging was studied. For the
emerging eigenvalues the asymptotics expansions were obtained as
the lengths of the window close to critical ones. The behaviour
of the associated eigenfunction was described as well.
Scattering for the system of two waveguides was considered in
\cite{ESTV}, \cite{Ku}. The case in which the Neumann condition
is imposed on the boundary instead of the Dirichlet one, was
studied in \cite{DK}. The existence of at least one isolated
eigenvalue was proven. In the paper \cite{BEK} the system of two
symmetric waveguides put in a magnetic field was considered. It
was shown that a magnetic field can eliminate the influence of
the window presence, namely, for sufficiently small window the
system has no bound states. At the same time, the system has a
bound state if the window is large enough.

\begin{figure}[t]
\begin{center}
\noindent
\includegraphics[width=13 true cm, height=6 true cm]{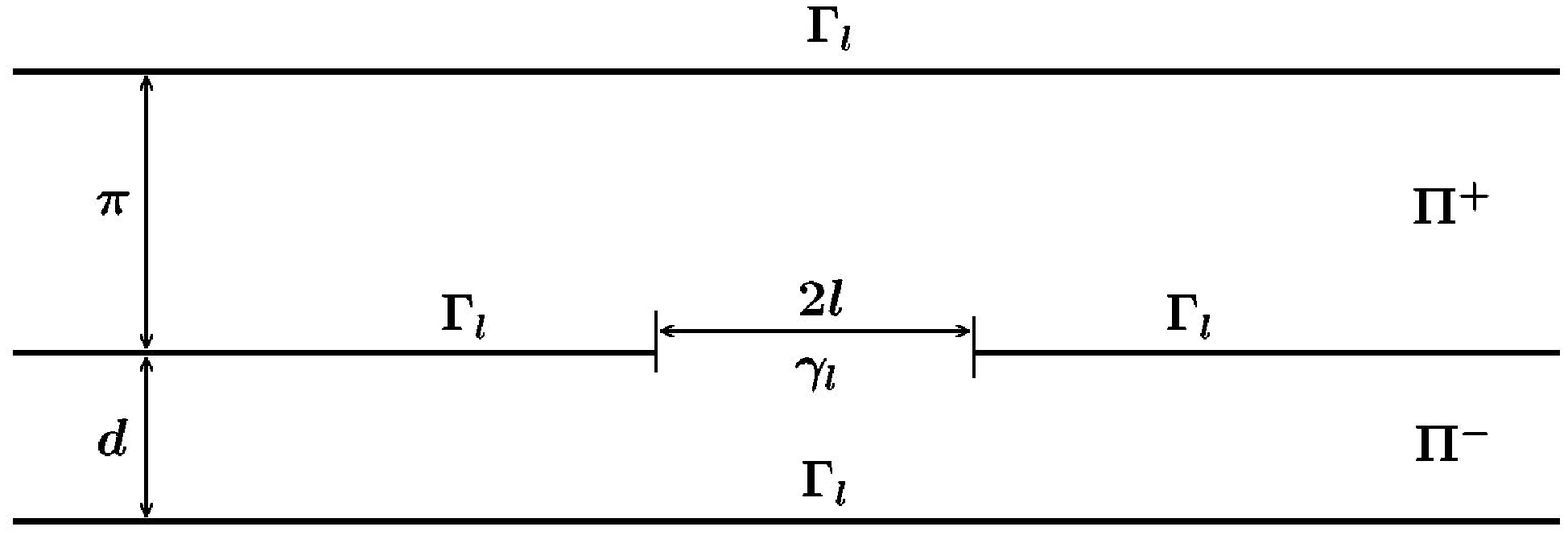}%

\noindent Figure.
\end{center}
\end{figure}

In the present paper we consider a pair of nonsymmetric
waveguides coupled by a finite window. The first part of the
work is devoted to the studying of the eigenvalue appearing
under the length of the window increasing. As it was mentioned,
the eigenvalues emerge when the length of the window passes
through some critical values. In the paper we give the criterion
of the ''criticality'' for a given value of the length. We also
obtain the asymptotics expansions for the emerging eigenvalues
and describe the behaviour of the associated eigenfunctions.
Moreover, we improve the two-sided estimates obtained in
\cite{ESTV}.

In the second part of the work we study the behaviour of the
discrete spectrum as the window widens. We obtain the
asymptotics expansions for the eigenvalues in this case. Under
the window widening the shift of the essential spectrum occurs
in the limits. We describe how this happens.

\section{Statement of the problem and formulation of the
results}

Let $x=(x_1, x_2)$ be Cartesian coordinates, $\Pi^+:=\{x:
0<x_2<\pi\}$, $\Pi^-:=\{x: -d<x_2<0\}$. The width $d$ of the
strip $\Pi^-$ is assumed to be not exceeding the width of the
strip $\Pi^+$. In the axis $x_2=0$ we select an interval
$\gamma_l$ of length $2l$ centered at zero which will be called
window in what follows. The union of the strips $\Pi^-$ and
$\Pi^+$ and the interval $\gamma_l$ is denoted by $\Pi$, i.e.,
the set $\Pi$ are strips $\Pi^+$ and $\Pi^-$ coupled by the
window $\gamma_l$. The boundary of the domain $\Pi$ is indicated
as $\Gamma_l$ (cf. figure).

The main object of our study is the spectrum of the operator
${H}_l:=-\Delta^\mathcal{(D)}_l$ in $L_2(\Pi)$, where
$\Delta^\mathcal{(D)}_l$ is the Friedrich's extension of the
Laplace operator from the set $C^\infty_0(\Pi)$. The essential
spectrum of the operator ${H}_l$ coincides with the real
semi-axis $[1,+\infty)$. For $l=0$ (i.e., in the case
$\gamma_l=\emptyset$, $\Pi=\Pi^+\cup\Pi^-$) it is obvious, while
the essential spectrum of the operators ${H}_0$ and ${H}_l$,
$l>0$, are same. The proof of this fact reproduces word for word
the proof of Theorem~2.1 in \cite{BEK} and based on the ideas of
the work \cite{B}. One just needs to take into account that the
domain $\Pi$ possesses the cone property (see definition in
\cite[Item 4.3, Ch. IV]{Ad}), thus by Rellich-Kondrashov theorem
\cite[Theorem 6.2, Ch. VI]{Ad} the embedding $W_2^1(Q\cap\Pi)\to
L_2(Q\cap\Pi)$ is compact  for any bounded subdomain
$Q\subset\Pi$ with smooth boundary.

As it has been mentioned in Introduction, the presence of the
window ($l>0$) gives rise to a non-empty discrete part of the
spectrum of the operator ${H}_l$, i.e., to the existence of the
isolated eigenvalues $\lambda_m(l)$, $m\geqslant1$. We take
these eigenvalues in ascending order with the multiplicity taken
into account.

In \cite{ESTV} the following statement was proved.

\begin{lemma}\label{lm1.1}
For any $l>0$ the operator ${H}_l$ has a non-empty discrete
spectrum consisting of finitely many eigenvalues. There exists
an infinite set of critical values $0=l_1<l_2<\ldots<l_n<\ldots$
of length of the window $\gamma_l$, such that as
$l\in(l_n,l_{n+1}]$ the operator ${H}_l$ has exactly $n$
isolated eigenvalues. These eigenvalues are non-increasing
functions on $l$ and satisfy two-sided estimates:
\begin{equation}\label{1.1}
\Lambda_{m-1}(l)
\leqslant\lambda_m(l)\leqslant\Lambda_{m}(l),\quad m\geqslant
1,\quad l>l_m,
\end{equation}
where
\begin{equation*}
\Lambda_m(l):=\frac{\pi^2}{(\pi+d)^2}+\frac{\pi^2 m^2}{4l^2}.
\end{equation*}
The number of eigenvalues $\lambda_m(l)$ meets the inequalities
\begin{equation*}
\left[\frac{2l}{\pi}\sqrt{1-\frac{\pi^2}{(\pi+d)^2}}\right]\leqslant
\mathrm{card}\,(\sigma_{disc}({H}_l))\leqslant
\left[\frac{2l}{\pi}\sqrt{1-\frac{\pi^2}{(\pi+d)^2}}\right]+1,
\end{equation*}
where $[\cdot]$ indicates an integer part.
\end{lemma}

Throughout the work by $W_2^1(\Omega,\gamma)$ we indicate the
completion in the norm of $W_2^1(\Omega)$ of the set of
functions from $C^\infty(\overline{\Omega})$ having compact
support and vanishing in a neighbourhood of the set $\gamma$. We
also set $\Pi_a:=\{x: |x_1|<a\}\cap\Pi$,
$\Gamma_l^a:=\Gamma_l\cap\partial\Pi_a$. By $\Xi$ we denote the
set of all bounded subdomains $Q\subset\Pi$ with smooth boundary
separated from the edges of the window $\gamma_l$ by a positive
distance. The case $\partial Q\cap\partial\Pi\not=\emptyset$ is
not excluded.

Let us formulate the main results of the present work.

\begin{theorem}\label{th1.1}
The statements are valid:
\begin{enumerate}
\item\label{th1.1.it1}
The eigenvalues $\lambda_m(l)$ of the operator ${H}_l$ are
continuous on $l$, simple and satisfy the estimates
\begin{equation}\label{1.10}
\Lambda_{m-1}(l)
<\lambda_m(l)<\Lambda_{m}(l),\quad
m\geqslant 1, \quad l>l_m.
\end{equation}
The associated eigenfunctions are even on $x_1$ for odd $m$, and
odd on $x_1$ for even $m$.

\item\label{th1.1.it3}
The length $l=l_n$ is critical, if and only if  a boundary value
problem
\begin{equation}\label{1.2}
-\Delta\phi_n=\phi_n,\quad x\in\Pi,\qquad \phi_n=0,\quad
x\in\Gamma_l,
\end{equation}
has a bounded solution belonging to $W_2^1(\Pi_a,\Gamma_l^a)$
for any $a>0$ and being even on $x_2$ in the case $d=\pi$, and
obeying an asymptotics representation
\begin{equation}\label{1.3}
\phi_n(x)=\sqrt{\frac{2}{\pi}}\sin x_2+\mathcal{O}(\mathrm{
e}^{-\sqrt{3}x_1}),\qquad x_1\to+\infty,\quad x_2\in(0,\pi).
\end{equation}
In the case such solution exists, it is unique and even on $x_1$
for odd $n$ and odd on $x_1$ for even $n$.

\item\label{th1.1.it4}
The asymptotics expansion of the eigenvalue $\lambda_n(l)$,
$n\geqslant2$, as $l\to l_n+0$ is as follows:
\begin{align}
&\lambda_n(l)=1-\mu_n^2(l-l_n)^2+\mathcal{O}\left((l-l_n)^3\right),
\label{1.4b}
\\
&\mu_n=\frac{1}{l_n}\int\limits_\Pi\left|\frac{\partial\phi_n}{\partial
x_1 }\right|^2\,dx\hphantom{2}\quad \text{as} \quad
d<\pi,\label{1.4}
\\
&
\mu_n=\frac{1}{2l_n}\int\limits_\Pi\left|\frac{\partial\phi_n}{\partial
x_1 }\right|^2\,dx\quad \text{as} \quad d=\pi.\label{1.4a}
\end{align}
The associated eigenfunction can be chosen such that it meets
the asymptotics representation
\begin{equation}\label{1.5}
\psi_n(x)=\sqrt{\frac{2}{\pi}}
\mathrm{e}^{-\sqrt{1-\lambda_n(l)}|x_1|}\sin
x_2+\mathcal{O}(\mathrm{e}^{-\sqrt{3-\lambda_n(l)}x_1}),\quad
x_1\to+\infty,\quad x_2\in(0,\pi).
\end{equation}
At the same time for any $R>0$ the equality
\begin{equation}\label{1.6}
\psi_n(x)=\phi_n(x)+\mathcal{O}\left((l-l_n)^{1/2}\right) \quad
\text{in norm}\quad W_2^1(\Pi_R),
\end{equation}
holds true.
\end{enumerate}
\end{theorem}

\begin{remark}\label{rm1.1}
In Item~\ref{th1.1.it3} of Theorem~\ref{th1.1} a solution to the
boundary value problem~\ref{1.2} is regarded in a generalized
sense. Namely, a solution is a function belonging to the space
$W_2^1(\Pi_a,\Gamma_l^a)$ for each $a>0$, and solving an
integral equation:
\begin{equation}\label{1.11}
\left(\nabla_x\phi_n,\nabla_x
\zeta\right)_{L_2(\Pi)}=(f,\zeta)_{L_2(\Pi)}
\end{equation}
for each function $\zeta\in C^\infty_0(\Pi)$. In accordance with
the theorems on improving smoothness of solutions to elliptic
problems \cite[Ch. 4, \S 2]{Ld}, the function $\phi_n$ belongs
to $C^\infty(\overline{Q})$ for each $Q\in\Xi$. This is why the
asymptotics (\ref{1.3}) should be understood in the usual sense.
In what follows all the boundary value problems are treated in
the sense of an integral equation similar to (\ref{1.11}).
Moreover, due to the theorems on improving smoothness solutions
to all boundary value problems posed in unbounded domains are
infinitely differentiable functions as the absolute value of
$x_1$ is large enough. This allows us to understand all the
statements on behaviour of these solutions at infinity in the
usual sense.
\end{remark}

\begin{remark}\label{rm1.2}
The function $\phi_n$ in Item~\ref{th1.1.it3} of
Theorem~\ref{th1.1} is supposed to be even on $x_2$ as $d=\pi$.
Such a restriction is needed to exclude from consideration the
function $\sqrt{2/\pi}\sin x_2$ which is a bounded solution to
the problem (\ref{1.2}) and satisfy the asymptotics
representation (\ref{1.3}) for all $l\geqslant 0$ in the case
$d=\pi$. In the case $d<\pi$ a solution similar to
$\sqrt{2/\pi}\sin x_2$ is absent and the requirement of being
even on $x_2$ is not introduced.
\end{remark}

\begin{remark}\label{rm1.3}
It should be noted that Item~\ref{th1.1.it4} of
Theorem~\ref{th1.1} was proved in \cite{BEG} for the case of
symmetric strips ($d=\pi$).
\end{remark}

\begin{theorem}\label{th1.2}
The following statements are valid:
\begin{enumerate}
\item\label{th1.2.it1}
The eigenvalues $\lambda_m(l)$ have the following asymptotics
expansion as $l\to+\infty$:
\begin{equation}\label{1.7}
\lambda_m(l)=\frac{\pi^2}{(\pi+d)^2}+\frac{\pi^2
m^2}{4l^2}+\mathcal{O}(l^{-3}).
\end{equation}
\item\label{th1.2.it2}
Each point of semi-interval
$\left[\frac{\pi^2}{(\pi+d)^2},1\right)$ is the accumulation
point for the eigenvalues  $\lambda_m(l)$ as $l\to+\infty$,
namely, for each point
$\xi\in\left[\frac{\pi^2}{(\pi+d)^2},1\right)$ there exists a
sequence of indexes $m=m(l,\xi)$ tending to infinity as
$l\to+\infty$, such that the convergence
\begin{equation*}
\lambda_{m(l,\xi)}\to\xi \quad\text{as}\quad l\to+\infty
\end{equation*}
holds true.
\end{enumerate}
\end{theorem}

Let us discuss the results of the work. Theorem~\ref{th1.1} is
devoted mostly to phenomenon of new eigenvalues of the operator
$H_l$ emerging as the window $\gamma_l$ widens. The first item
of the theorem improves the estimate (\ref{1.1}), the second one
provides the criterion determining the critical values of the
window $\gamma_l$. As it follows from the third item of
Theorem~\ref{th1.1}, new eigenvalues emerge from the threshold
of the essential spectrum of the operator ${H}_l$ and have the
asymptotics expansion (\ref{1.4b})--(\ref{1.4a}). The leading
term of this expansion is nonzero. This fact follows easily from
the formula for $\mu_n$ and the boundary value problem for
$\phi_n$. Formulas (\ref{1.4}) and (\ref{1.4a}) imply that the
coefficient $\mu_n$ is discontinuous as $d\to\pi$. Earlier
similar phenomenon for the eigenvalue $\lambda_1(l)$ as $l$ is
small enough was found formally in \cite{P}.

The second part of the results given in Theorem~\ref{th1.2}
describe the behaviour of the spectrum of the operator ${H}_l$
as the length of the window increases. As it follows from the
first item of Theorem~\ref{th1.2}, all the eigenvalues
$\lambda_m(l)$ tend to the threshold of the essential spectrum
of the ''limiting'' operator, coinciding up to a quantity of
order $\mathcal{O}(l^{-3})$ with the right end-points of the
intervals from Item~\ref{th1.1.it1} of Theorem~\ref{th1.1}. We
stress that the estimate for the error term in (\ref{1.7}) is
not uniform on $m$. We also note that the leading term in the
asymptotics expansion (\ref{1.7}) is independent on $d$ in
contrast to the formula (\ref{1.4}) where this parameter plays a
crucial role.

As the length of the window increases, it is appropriate to
compare the spectra of the original operator $H_l$ and a
''limiting'' operator $H_*:=-\Delta^{(\mathcal{D})}_*$, where
$\Delta^{(\mathcal{D})}_*$ is the Friedrich's extension of the
Laplace operator from a set  $C^\infty_0(\Pi^*)$, $\Pi^*:=\{x:
-d<x_2<\pi\}\setminus\{x: x_1\geqslant0, x_2=0\}$. This
''limiting'' operator appears if in the original problem one
makes a shift $x_1 \mapsto x_1-l$ and pass formally to the limit
as $l\to+\infty$. The spectrum of the operator $H_*$ consists of
its essential part only and coincides with the semi-axis
$\left[\frac{\pi^2}{(\pi+d)^2},+\infty\right)$. In order to
prove this fact one just needs to estimate the threshold of the
essential spectrum of the operator $H_*$ both from above and
below by bracketing \cite[Ch. 13, \S 15]{RS}, introducing in the
domain $\Pi^*$ an additional boundary $\{x: x_1=0, -d<x_2<\pi\}$
and imposing Dirichlet or Neumann condition on it.

The second item of Theorem~\ref{th1.2} describes how the shift
of the essential spectrum occurs as $l\to+\infty$: each point of
the semi-interval which is the shift of the essential spectrum
in the limit is an accumulation point as $l\to+\infty$ for the
eigenvalues $\lambda_m(l)$ whose indexes increases unboundedly
together with $l$.

Let us describe briefly the structure of the present work. In
the next section we prove Item~\ref{th1.1.it1} of
Theorem~\ref{th1.1} as well as the convergence of the
eigenvalues to the threshold of the essential spectrum as the
length of the window tends to a critical size. The third section
is devoted to the studying behaviour of the resolvent as the
spectral parameter tends to the threshold of the essential
spectrum. Basing on the results of the third section, in the
fourth one we prove Items~\ref{th1.1.it3} and \ref{th1.1.it4} of
Theorem~\ref{th1.1}. The proof of Theorem~\ref{th1.2} is adduced
in the last section.

\section{Estimates, continuity and convergence of
{eigen\-values}}

The present section is devoted to the proof of
Item~\ref{th1.1.it1} of Theorem~\ref{th1.1}. We will also prove
the convergence of the eigenvalues  $\lambda_n(l)$ to the
threshold of the essential spectrum as $l\to l_n+0$.

\begin{lemma}\label{lm2.1}
The eigenvalues $\lambda_m(l)$ are continuous on $l$. As $l\to
l_n+0$ the convergence $\lambda_n\to1-0$ holds true.
\end{lemma}
\begin{proof}
According to Lemma~\ref{lm1.1}, the operator ${H}_l$ is lower
semibounded and its lower bound is $\frac{\pi^2}{(\pi+d)^2}$.
Therefore, for each value of $l$ there exists a bounded inverse
operator ${H}_l^{-1}: L_2(\Pi)\to L_2(\Pi)$. The functions
$\lambda_m^{-1}(l)$ are isolated eigenvalues of the operator
${H}_l^{-1}$. Let us prove that they are continuous on $l$. Let
$l_*$ be a given length of the window $\gamma_l$ and
$\lambda_m(l_*)$ is an isolated eigenvalue of the operator
${H}_{l_*}$. 
The eigenvalue $\lambda_m(l)$ is obviously to be an eigenvalue
of the boundary value problem
\begin{equation}\label{2.2}
-\Delta\psi=\lambda\psi,\quad x\in\Pi, \qquad \psi=0,\quad
x\in\Gamma_l.
\end{equation}
We remind that a solution to this boundary value problem is
regarded in the generalized sense (see Remark~\ref{rm1.1}). Due
to Theorem~4.6.8 from \cite{BEH} it guarantees the belonging of
a generalized solution to the domain of the operator ${H}_l$, if
its solution is an element of $L_2(\Pi)$.

Let $\chi_1(x_1)$ be an infinitely differentiable cut-off odd
function which equals minus one as
$x_1\in[-l_*-\varepsilon_0,-l_*+\varepsilon_0]$, is one as
$x_1\in[l_*-\varepsilon_0,l_*+\varepsilon_0]$, and vanishes as
$x_1\in(-\infty,-l_*-2\varepsilon_0]\cup[-l_*+2\varepsilon_0,
l_*-2\varepsilon_0]\cup [l_*+2\varepsilon_0,+\infty)$, where
$\varepsilon_0$ is a small fixed number. In the problem
(\ref{2.2}) we make a change of variables
\begin{equation}\label{2.0}
y_1=x_1-\varepsilon\chi_1(x_1),\quad y_2=x_2,\quad
\varepsilon=l-l_*,\quad
\varepsilon\in[-\varepsilon_0,\varepsilon_0].
\end{equation}
Such change, as it can be checked easily, leads us to a new
boundary value problem:
\begin{equation}\label{2.1}
\begin{aligned}
-&(\Delta_y+\varepsilon L_\varepsilon)\psi=\lambda\psi,\quad
y\in\Pi,\qquad \psi=0,\quad y\in\Gamma_{l_*},
\\
&L_\varepsilon=A_{11}(y_1,\varepsilon)\frac{\partial^2}{\partial
y_1^2}+A_1(y_1,\varepsilon)\frac{\partial}{\partial y_1},
\\
&A_{11}(y_1,\varepsilon)=-2\chi'_1\big(x_1(y_1,\varepsilon)\big)+
\varepsilon\left(\chi'_1\big(x_1(y_1,\varepsilon)\big)\right)^2,
\\
&A_1(y_1,\varepsilon)=-\chi''_1\big(x_1(y_1,\varepsilon)\big).
\end{aligned}
\end{equation}
Therefore, the function $\lambda_m^{-1}(l)$ is an eigenvalue of
the operator $({H}_{l_*}+\varepsilon L_\varepsilon)^{-1}:
L_2(\Pi)\to L_2(\Pi)$. This operator is well-defined and
bounded. Indeed, the operator ${H}_{l_*}^{-1}$ is a bounded
operator from $L_2(\Pi)$ into $W_2^1(\Pi)$ and $W_2^2(Q)$ for
each $Q\in\Xi$. The boundedness of the operator ${H}_{l_*}^{-1}:
L_2(\Pi)\to W_2^1(\Pi)$ is obvious while the boundedness of the
operator ${H}_{l_*}^{-1}: L_2(\Pi)\to W_2^2(Q)$ follows from
theorems on improving smoothness of solutions to elliptic
boundary value problems \cite[Ch. 4, \S 2]{Ld}. Taking into
account these facts as well as boundedness and compactness of
supports of the coefficients of the operator $L_\varepsilon$, we
conclude that the operator ${H}_{l_*}^{-1}L_\varepsilon$ is
bounded uniformly on $\varepsilon$ as an operator from
$L_2(\Pi)$ into $L_2(\Pi)$. Thus, the operator
$({H}_{l_*}+\varepsilon L_\varepsilon)^{-1}: L_2(\Pi)\to
L_2(\Pi)$ is well-defined for sufficiently small $\e$. It is
easy to check that it is determined by the formula
$({H}_{l_*}+\varepsilon L_\varepsilon)^{-1}=(I+\varepsilon
{H}_{l_*}^{-1}L_\varepsilon)^{-1}{H}_{l_*}^{-1}$. The last
representation proves also the convergence of the operator
$({H}_{l_*}+\varepsilon L_\varepsilon)^{-1}$ to ${H}_{l_*}^{-1}$
in the operator norm as $\varepsilon\to0$. From \cite[Ch. 4, \S
2.6, Theorem 2.23]{K} it follows that the operator
$(H_{l_*}+\varepsilon L_\varepsilon)^{-1}$ converge to
$H_{l_*}^{-1}$ in a generalize sense as well. In its turn, due
to \cite[Ch. 4, \S 3.5]{K} it implies the convergence
$\lambda_m^{-1}(l)\to\lambda_m^{-1}(l_*)$ as $l\to l_*$, what
proves the needed continuity of the eigenvalues of the operator
${H}_l$.

Let us prove the convergence of the eigenvalues $\lambda_n(l)$
to the threshold of the essential spectrum as  $l\to l_n-0$. The
convergence $\lambda_1(l)\xrightarrow[l\to+0]{}1$ follows from
\cite[Theorem 2.1]{EV2}. The eigenvalues $\lambda_n(l)$ are
monotonically nondecreasing functions on $l$ bounded from above
by one. This yields the existence of the limits
$c_n=\lim\limits_{l\to l_n+0}\lambda_n(l)$. Suppose that one of
these limits is strictly less than one. Then the number $c_n$ is
an eigenvalue of the operator ${H}_l$ as $l=l_n$ (see the proof
of the continuity of the eigenvalues on $l$ adduced above).
Hence, as $l=l_n$ the operator $H_l$ has $n$ isolated
eigenvalues what contradicts to Lemma~\ref{lm1.1}.
\end{proof}

\begin{lemma}\label{lm2.2}
The Item~\ref{th1.1.it1} of Theorem~\ref{th1.1} is valid.
\end{lemma}
\begin{proof}
The continuity of the eigenvalues was proved in the previous
lemma. The simplicity of the eigenvalues $\lambda_m$ is surely
to be an implication of the estimates~\ref{1.10}. Let us prove
the latter. According to minimax principle the eigenvalues of
the operator ${H}_l$ are given by the formulas
\begin{equation}\label{2.3}
\lambda_m(l):=\inf\limits_{\genfrac{}{}{0 pt}{}{u\in
W_2^1(\Pi,\partial\Pi), u\not=0,} { \genfrac{}{}{0
pt}{}{(u,\psi_j)_{L_2(\Pi)}=0,}{j=1,\ldots,m-1} } }
\frac{\|\nabla u \|^2_{L_2(\Pi)}}{\|u\|^2_{L_2(\Pi)}},
\end{equation}
where, we remind, $\psi_j$ are the eigenfunctions associated
with $\lambda_j(l)$. We introduce the functions
\begin{equation*}
u_j(x)=\left\{
\begin{aligned}
&{\frac{\sqrt{2}}{\sqrt{\pi+d}}}\sin\frac{\pi}{\pi+d}(x_2-\pi)
\sin\frac{\pi j}{2l}(x_1+l),\quad&& x\in\Pi_l,
\\
&\hphantom{\sqrt{\frac{2}{\pi+d}}\sin\frac{\pi}{\pi+d}} 0,
&&x\not\in\Pi_l.
\end{aligned} \right.
\end{equation*}
Clear, the functions $u_j$ belong to the space
$W_2^1(\Pi,\partial\Pi)$. Let us prove the right-hand side of
the estimates (\ref{1.10}). Suppose the opposite, namely, let
for some $l$ and $m$ the equality $\lambda_m(l)=\Lambda_m(l)$ is
true. The functions $u_j$ are linear independent, this is why in
the linear space spanned on the functions $u_j$, $j=1,\ldots,m$,
there exists a nonzero function $u=\sum_{j=1}^m\alpha_j u_j$
being orthogonal in  $L_2(\Pi)$ to each function $\psi_i$,
$i=1,\ldots,m-1$. By (\ref{2.3}) we have
\begin{equation}\label{2.5}
\lambda_m(l)\leqslant\frac{\|\nabla
u\|^2_{L_2(\Pi)}}{\|u\|^2_{L_2(\Pi)}}=\frac{\sum_{j=1}^m
\alpha_j^2\Lambda_j}{\sum_{j=1}^m \alpha_j^2}.
\end{equation}
The fraction in the right-hand side of this relation does not
exceed $\Lambda_m(l)$. The equality
\begin{equation*}
\frac{\sum_{j=1}^m \alpha_j^2\Lambda_j}{\sum_{j=1}^m
\alpha_j^2}=\Lambda_m(l)
\end{equation*}
is possible only in the case $\alpha_m\not=0$, $\alpha_j=0$,
$j=1,\ldots,m-1$. In this case the function $u_m$ is an
eigenfunction of the operator ${H}_l$ associated with the
eigenvalue $\lambda_m(l)$. This contradicts to the fact that all
the eigenfunctions of the operators ${H}_l$ belong to
$C^\infty(\Pi)$. Thus, at least one of numbers $\alpha_j$,
$j=1,\ldots,m-1$, is nonzero, what by (\ref{2.5}) yields the
estimate for $\lambda_m(l)$:
\begin{equation*}
\lambda_m(l)\leqslant\frac{\|\nabla
u\|^2_{L_2(\Pi)}}{\|u\|^2_{L_2(\Pi)}}=\frac{\sum_{j=1}^m
\alpha_j^2\Lambda_j}{\sum_{j=1}^m \alpha_j^2}<\Lambda_m(l).
\end{equation*}
This contradicts to the original assumption that
$\lambda_m(l)=\Lambda_m(l)$.

We proceed to the proof of the left-hand side of the estimates
(\ref{1.10}). Let the operator ${H}_l$ has $n$ eigenvalues, what
due to Lemma~\ref{lm2.1} implies that $\Lambda_j(l)<1$,
$j=0,\ldots,n-1$. Let $\delta>0$ be some small number. Through
the points $(-l+\delta,0)$ and $(l-\delta,0)$ we pass the
segments being parallel to the axis $x_1=0$ and dissecting $\Pi$
into three disjoint parts. Isolated eigenvalues of the Laplacian
in $\Pi$ subject to Dirichlet condition on $\Gamma_l$ and
Neumann condition on the segments introduced estimate the
eigenvalues $\lambda_m(l)$ from below. The essential spectrum of
such operator coincides with real semi-axis $[1,+\infty)$, what
can be established in same way as the equality
$\sigma_{ess}\,({{H}_l})=[1,+\infty)$. The discrete spectrum of
this operator is a union $\sigma_1\cup\sigma_2$, where
$\sigma_1$ is a set of the eigenvalues of the operator $S_1$
that are less than one. Here the operator $S_1$ is the Laplacian
in an rectangle $\{x: |x_1|<l-\delta, -d<x_2<\pi\}$ subject to
Neumann condition on the lateral sides and to Dirichlet one on
the upper and lower sides. The set $\sigma_2$ is the discrete
spectrum of the Laplacian in the semi-strip $\Pi\cap\{x:
x_1>l-\delta\}$ subject to Neumann condition on $\{x:
x_1=l-\delta, -d<x_2<\pi\}$ and to Dirichlet condition on the
remaining part of the boundary. We denote this operator by
$S_2$. For sufficiently small $\delta$ the eigenvalues forming
$\sigma_1$ are the functions $\Lambda_j(l-\delta)$,
$j=0,\ldots,n-1$. Each eigenfunction of the operator $S_2$ can
be continued through the boundary $\{x: x_1=l-\delta,
-d<x_2<\pi\}$ in the odd way on $x_1$. The function obtained in
this way is the eigenfunction of the operator ${H}_{\delta}$ (up
to the change $x_1\mapsto x_1-l+\delta$). Therefore,
$\sigma_2\subseteq\sigma_{disc}({H}_\delta)$. In accordance with
Lemma~\ref{lm1.1}, for sufficiently small $\delta$ the discrete
spectrum of the operator ${H}_\delta$ consists of the only
eigenvalue converging to one as $\delta\to0$. We choose
$\delta>0$ such that this eigenvalue is greater than each of the
functions $\Lambda_j(l-\delta)$, $j=0,\ldots,n-1$. Therefore,
due to bracketing \cite[Ch. 13, \S 15]{RS} we can write
$\Lambda_{j-1}(l)<\Lambda_{j-1}(l-\delta)\leqslant\lambda_j(l)$,
$j=1,\ldots,n$, what completes the proof of the needed
estimates.

In conclusion let us prove the parity of the eigenfunctions of
the operator ${H}_l$. The set $\Pi$ being symmetric on $x_1$,
all the eigenfunctions of the operator ${H}_l$ can be chosen as
being odd or even on $x_1$. The simplicity of the eigenvalues
$\lambda_m(l)$ means that the eigenfunction of a certain parity
is associated with each of these eigenvalues. The even
eigenfunctions satisfy the Neumann condition as $x_1=0$, while
the odd ones meet the Dirichlet condition. Moreover, the
operator ${H}_l$ is an orthogonal sum of the operators ${H}_l^+$
and ${H}_l^-$ those are, respectively, restrictions of ${H}_l$
on even and odd on $x_1$ functions from the domain of the
operator ${H}_l$. Completely by analogy with how in \cite[\S
2]{ESTV} the estimates (\ref{1.1}) were obtained, one can easily
show that the isolated eigenvalues of the operator ${H}_l^+$
satisfy the estimates (\ref{1.1}) for odd $m$, while the ones of
the operator $H_l^-$ meet the estimates (\ref{1.1}) with even
$m$. This proves the needed parity of the eigenfunctions of the
operator ${H}_l$, if one takes into account that
$\sigma_{disc}({H}_l)=\sigma_{disc}({H}_l^+)\cup
\sigma_{disc}({H}_l^-)$.
\end{proof}

\section{The behaviour of the resolvent of
the operator ${H}_l$ in a vicinity of the threshold of the
essential spectrum}

This section is devoted to the studying the behaviour of the
operator $({H}_l-\lambda)^{-1}$ as $\lambda$ close to one. The
results of this section is the basis for the proof of
Items~\ref{th1.1.it3},~\ref{th1.1.it4} of Theorem~\ref{th1.1}.

In studying the operator $({H}_l-\lambda)^{-1}$ we employ the
same approach as that used in \cite{BEG}, \cite{BE} for the case
of symmetric strips $d=\pi$. We study the dependence on $k$ of a
solution to the boundary value problem
\begin{equation}\label{3.1}
-\Delta u=(1-k^2)u+f,\quad x\in\Pi,\qquad u=0,\quad
x\in\partial\Pi,
\end{equation}
which behaves as follows
\begin{equation}\label{3.2}
\begin{aligned}
&u(x,k)=c_{\pm}(k)\mathrm{e}^{-k|x_1|}\sin
x_2+\mathcal{O}\left(\mathrm{e}^{-\sqrt{3+k^2}|x_1|}\right),&&
x_2\in(0,\pi),
\\
&u(x,k)=\widetilde
c_{\pm}(k)\mathrm{e}^{-\sqrt{1-\frac{\pi^2}{d^2}+k^2}|x_1|}\sin
x_2+\mathcal{O}\left(\mathrm{
e}^{-\sqrt{4-\frac{\pi^2}{d^2}+k^2}|x_1|}\right), &&
x_2\in(-d,0),
\end{aligned}
\end{equation}
as $x_1\to\pm\infty$. Here the function $f$ is an element of
$L_2(\Pi)$ whose support lies inside $\Pi_a$, $a>l$, $c_\pm(k)$,
$\widetilde c_\pm(k)$ are some constants. In the case $d=\pi$ in
the latter of the asymptotics representations (\ref{3.2}) we set
$\sqrt{1-\frac{\pi^2}{d^2}+k^2}=k$. The parameter $k$ is
supposed to belong to a small neighbourhood of the zero in the
complex plane. We denote this neighbourhood by ${B}$. We note
that a solution to the boundary value problem (\ref{3.1}),
(\ref{3.2}) decays exponentially as $\mathrm{Re}\,k>0$, and,
therefore, is an element of $L_2(\Pi)$ in this case. In view of
Remark~\ref{rm1.1} and \cite[Theorem 4.6.8]{BEH} it implies the
belonging of this solution to the domain of the operator
${H}_l$, i.e., the function $u$ coincide with
$(H_l-1+k^2)^{-1}f$ (of course, if the operator $(H_l-1+k^2)$ is
invertible). This is why the linear mapping $f\mapsto u$ defined
by the boundary value problem (\ref{3.1}), (\ref{3.2}) can be
regarded as an extension of the operator $(H_l-1+k^2)^{-1}$ on
$k$ in the domain $\mathrm{Re}\, k\leqslant0$. Such extension is
surely to widen the range of the operator $(H_l-1+k^2)^{-1}$ and
the range of the extension is not a subset of the space
$L_2(\Pi)$. At the same time we will show that in a certain
sense this extension is analytic on $k$ and the operator
$(H_l-1+k^2)^{-1}$ after extension is happened to be meromorphic
on $k$.

Let us introduce the notations. If $X$ and $Y$ are Banach
spaces, the symbol $\mathcal{L}(X,Y)$ indicate the set of all
linear bounded operator from $X$ into $Y$. The set of all
holomorphic (meromorphic) on $k\in{B}$ function whose values are
elements of $X$ is denoted by $\mathcal{H}(X)$
($\mathcal{M}(X)$). We also set
$\mathcal{H}(X,Y):=\mathcal{H}(\mathcal{L}(X,Y))$,
$\mathcal{M}(X,Y):=\mathcal{M}(\mathcal{L}(X,Y))$.

In order to study the boundary value problem (\ref{3.1}) we
employ the scheme borrowed from \cite[Ch. 16, \S 4]{SP}. Let $g$
be some function from $L_2(\Pi_a)$ continued by zero in
$\Pi\setminus\overline{\Pi}_a$. We consider the boundary value
problems:
\begin{equation}\label{3.3}
-\Delta v_i=(1-k^2)v_i+g,\quad x\in\Omega_i, \qquad v_i=0,\quad
x\in\partial\Omega_i,\quad i=1,\ldots,4,
\end{equation}
where $\Omega_1:=\Pi^+\cap\{x: x_1>0\}$,
$\Omega_2:=\Pi^-\cap\{x: x_1>0\}$,  $\Omega_3:=\Pi^+\cap\{x:
x_1<0\}$, $\Omega_4:=\Pi^-\cap\{x: x_1<0\}$. The problems
(\ref{3.3}) are easily solved by separation of variables:
\begin{align}
&v_i(x,k)=\sum\limits_{j=1}^\infty\int\limits_{\Omega_i}
G^i_{\!j}(x,t,k)g(t)\,dt,\label{3.4}
\\
&G_j^1(x,t,k):=\frac{1}{\pi s_j^+}\left(\mathrm{
e}^{-s_j^+|x_1-t_1|}- \mathrm{e}^{-s_j^+(x_1+t_1)}\right)\sin j
x_2\sin j t_2,\nonumber
\\
&G_j^2(x,t,k):=\frac{1}{s_j^- d}\left(\mathrm{
e}^{-s_j^-|x_1-t_1|}- \mathrm{e}^{-s_j^-(x_1+t_1)}\right)\sin
\frac{\pi j}{d} x_2\sin \frac{\pi j}{d} t_2,\nonumber
\\
&G_j^3(x,t,k):=\frac{1}{\pi s_j^+}\left(\mathrm{
e}^{-s_j^+|x_1-t_1|}- \mathrm{e}^{s_j^+(x_1+t_1)}\right)\sin j
x_2\sin j t_2, \nonumber
\\
&G_j^4(x,t,k):=\frac{1}{s_j^-
d}\left(\mathrm{e}^{-s_j^-|x_1-t_1|}- \mathrm{
e}^{s_j^-(x_1+t_1)}\right)\sin \frac{\pi j}{d} x_2\sin \frac{\pi
j}{d} t_2,\nonumber
\end{align}
where $s_1^+=k$, $s_j^+=\sqrt{j^2-1+k^2}$, $j\geqslant 2$,
$s_1^-=\sqrt{\frac{\pi^2}{d^2}-1+k^2}$ as $d<\pi$, $s_1^-=k$ as
$d=\pi$, $s_j^-=\sqrt{\frac{\pi^2j^2}{d^2}-1+k^2}$, $j\geqslant
2$. The functions $G_1^1$, $G_3^1$ at $k=0$ are defined by
continuity:
\begin{align*}
&G_1^1(x,t,0):=\frac{1}{\pi}(x_1+t_1-|x_1-t_1|)\sin x_2\sin t_2,
\\
&G_1^3(x,t,0):=-\frac{1}{\pi}(x_1+t_1+|x_1-t_1|)\sin x_2\sin
t_2,
\end{align*}
In the case $d=\pi$ the functions $G_1^2(x,t,0)$ and
$G_1^4(x,t,0)$ are defined in the same way. We denote
$\Omega_i^b:=\Omega_i\cap\Pi_b$.

\begin{lemma}\label{lm3.6}
Let $b>0$. The series (\ref{3.4}) converge in the norm of
$W_2^2(\Omega^b_i)$. The functions $v_i(x)$ meet the asymptotics
formulas (\ref{3.2}). Linear operators $T_i(k)$ defined by a
rule $T_i(k)g:=v_i$ are elements of
$\mathcal{L}(L_2(\Pi_a),W_2^2(\Omega_i^b))$. The belonging
$T_i(\cdot)\in \mathcal{H}(L_2(\Pi_a),W_2^2(\Omega_i^b))$ takes
place.
\end{lemma}

In proof of this lemma we will employ an auxiliary statement.
\begin{lemma}\label{lm3.7}
In the norm of $L_2(\Omega_i^a)$ the equality
\begin{align*}
&g(x):=\sum\limits_{j=1}^\infty g_j(x_1)\sin j x_2, \quad
g_j(x_1):=\frac{2}{\pi}\int\limits_{0}^\pi g(x)\sin j x_2\,
dx_2,\qquad i=1,3,
\\
&g(x):=\sum\limits_{j=1}^\infty g_j(x_1)\sin j x_2, \quad
g_j(x_1):=\frac{2}{d}\int\limits_{-d}^0 g(x)\sin \frac{\pi j}{d}
x_2\, dx_2,\qquad i=2,4,
\end{align*}
holds true.
\end{lemma}
\begin{proof}
We will give the proof for $\Omega_1^a$ only, in the other cases
the arguments are same. Since $g\in L_2(\Omega_i^a)$, by Fubini
theorem for almost each $x_1\in (0,a)$ we have $g(x_1,\cdot)\in
L_2(0,\pi)$. Therefore, the functions $g_j(x_1)$ are
well-defined for almost each $x_1\in (0,a)$ and belong to
$L_2(0,a)$ due to an estimate:
\begin{equation*}
\|g_j\|_{L_2(0,a)}\leqslant \|g\|_{L_2(\Pi)}.
\end{equation*}
We introduce the functions
\begin{equation*}
\mathcal{E}_N(x_1)=\int\limits_0^\pi
\left|g(x)-\sum\limits_{j=1}^N g_j(x_1)\sin j x_2\right|^2\,
dx_2.
\end{equation*}
The functions $\{\sin j x_2\}_{j=0}^\infty$ form basis in
$L_2(0,\pi)$, this is why the convergence
$\mathcal{E}_N(x_1)\xrightarrow[N\to\infty]{}0$ is valid  for
almost each $x_1\in (0,a)$. Using the definition of the
functions $\mathcal{E}_N$, one can check easily that
\begin{equation*}
0\leqslant \mathcal{E}_N(x_1)=\int\limits_0^\pi
|g(x)|^2\,dx_1-\frac{\pi}{2}\sum\limits_{j=1}^N|g_j(x_1)|^2\leqslant
\int\limits_0^\pi |g(x)|^2\,dx_1.
\end{equation*}
Therefore, nonnegative functions $\mathcal{E}_N$ are bounded
from above by an integrable over $[0,a]$ function uniformly on
$N$. Bearing in mind  this fact as well as the convergence of
the functions $\mathcal{E}_N$ to zero almost everywhere, in view
of Lebesgue bounded convergence theorem we conclude that
\begin{equation*}
\left\|g(x)-\sum\limits_{j=1}^N g_j(x_1)\sin j
x_2\right\|^2_{L_2(\Pi_a)}=\int\limits_{0}^a
\mathcal{E}_N(x_1)\,dx_1\xrightarrow[N\to\infty]{}0.
\end{equation*}
This completes the proof.
\end{proof}
\begin{proof}[Proof of Lemma~\ref{lm3.6}]
We will give the proof for $\Omega_1^a$ only, the other cases
are proved in the same way. We define the functions $g_j$ in
accordance with Lemma~\ref{lm3.7}. We indicate the terms of the
series (\ref{3.4}) as $V_j(x,k)$. By the definition of these
functions the estimates
\begin{equation*}
\left\| \sum\limits_{j=N}^M
V_j\right\|^2_{W_2^1(\Omega_1^b)}\leqslant C \sum\limits_{j=N}^M
\|g_j\|^2_{L_2(0,b)}
\end{equation*}
hold true for all $b>0$ with constant $C$ independent on $g$,
$M$ and $N$. The right-hand side in this inequality tends to
zero as $M,N\to\infty$ due to Lemma~\ref{lm3.7}. Thus, for all
$b>0$ the series (\ref{3.4}) converges in the norm of
$W_2^1(\Omega_1^b)$ to some function $v_1(x,k)$, which meets the
estimate
\begin{equation}\label{3.24}
\|v_1\|_{W_2^1(\Omega_1^b)}\leqslant C\|g\|_{L_2(\Pi_a)},
\end{equation}
where the constant $C$ is independent on $g$. The function
$v_1$, as one can check easily, is a generalized solution to the
boundary value problem (\ref{3.3}). Therefore, by theorems on
improving smoothness and the estimate (\ref{3.24}) the function
$v_1$ is an element of $W_2^2(\Omega_1^b)$ and an estimate
\begin{equation*}
\|v_1\|_{W_2^2(\Omega_1^b)}\leqslant C\|g\|_{L_2(\Pi_a)},
\end{equation*}
is valid, where the constant $C$ is independent on $g$. It
follows the belonging $T_1(k)\in
\mathcal{L}(L_2(\Pi_a),W_2^2(\Omega_1^b))$ for all $b>0$ and
each $k\in {B}$. By analogy with how the latter estimate for
$v_1$ has been obtained, one can deduce that
\begin{equation}\label{3.23}
\|V^N\|_{W_2^2(\Omega_1^b)}\leqslant C\|g^N\|_{L_2(\Pi_a)},
\end{equation}
where the constant is independent on $g^N$ and $N$,
\begin{equation*}
V^N(x,k):=v_1(x,k)-\sum\limits_{j=1}^N V_j(x,k),\quad
g^N(x):=g(x)-\sum\limits_{j=1}^N g_j(x_1)\sin jx_2.
\end{equation*}
The estimate (\ref{3.23}) and Lemma~\ref{lm3.7} yield the
convergence of the series (\ref{3.4}) in $W_2^2(\Omega_1^b)$.

For $x_1>a$ the functions $V_j$ are of the form
\begin{equation*}
V_j(x,k)=-\sum\limits_{j=1}^\infty \frac{2}{s_j^+\pi}\mathrm{
e}^{-s_j^+ x_1}\sin j x_2\int\limits_{\Omega_1^a}g(x)\sh s_j^+
x_1 \sin j x_2\,dx.
\end{equation*}
Thus, for $x_1>a$ the estimate
\begin{equation*}
|V_j(x,k)|\leqslant
C\mathrm{e}^{-s_j^+(x_1-a)}\|g\|_{L_2(\Pi_a)}
\end{equation*}
hold true, where $C$ is a some constant independent on $j$ and
$x_1$. In view of this inequality as $x_1\geqslant 2a$ the
function $(v_1-V_1)$ can be estimated as follows:
\begin{align*}
|v_1(x,k)-V_1(x,k)|&\leqslant
C\|g\|_{L_2(\Pi_a)}\sum\limits_{j=2}^{\infty}\mathrm{e}^{-s_j^+(x_1-a)}
\leqslant
\\
&\leqslant C\|g\|_{L_2(\Pi_a)}\mathrm{e}^{-s_2^+(x_1-a)}
\sum\limits_{j=2}^{\infty}\mathrm{e}^{-(s_j^+-s_2^+)(x_1-a)}\leqslant
\\
&\leqslant C\|g\|_{L_2(\Pi_a)}\mathrm{e}^{-s_2^+(x_1-a)}
\sum\limits_{j=2}^{\infty}\mathrm{e}^{-(s_j^+-s_2^+)a}\leqslant
 \widetilde C\|g\|_{L_2(\Pi_a)}\mathrm{e}^{-s_2^+x_1},
\end{align*}
where the constant $\widetilde C$ is independent on $x_1$. The
estimate obtained yields that as $x_1\geqslant 2a$,
$x_2\in[0,\pi]$ the series (\ref{3.4}) is a continuous on $x$
function and the asymptotics formula (\ref{3.2}) takes place for
the function $v_1$ as $x_1\to+\infty$, $x_2\in(0,\pi)$.

Clear, for each function $g$ we have $V_j(x,\cdot)\in
\mathcal{H}(W_2^2(\Omega_1^b))$. Since the series (\ref{3.4})
converges in $W_2^2(\Omega_1^b)$, by Weiestrass theorem the sum
of the series is holomorphic on $k$ in the norm of
$W_2^2(\Omega_1^b)$, i.e., for each function $g\in L_2(\Pi_a)$
and any $b>0$ we have $T_1(\cdot)g\in
\mathcal{H}(W_2^2(\Omega_1^b))$. Since the notions of being
holomorphic for bounded operator-valued functions in the sense
of weak, strong and uniform convergences are same (see, for
instance, \cite[Ch. 7, \S 1.1]{K}), we conclude that
$T_1(\cdot)\in \mathcal{H}(L_2(\Pi_a),W_2^2(\Omega_1^b))$ for
all $b>0$.
\end{proof}

\begin{remark}\label{rm3.1}
The functions $v_i$ being elements of the spaces
$W_2^2(\Omega_i^b)$, the equations (\ref{3.3}) take place not
only in the sense of the corresponding integral equality (see
remark~\ref{rm1.1}), but also as the equality of two functions
from $L_2(\Omega_i^b)$.
\end{remark}

We denote $v(x,k):=v_i(x,k)$, $x\in\Omega_i$. Let us consider
one more boundary value problem:
\begin{equation}\label{3.5}
\Delta w=\Delta v,\quad x\in\Pi_a, \qquad w=v,\quad
x\in\partial\Pi_a.
\end{equation}
The first derivatives of the function $v$ have discontinuities
on the boundaries of the sets $\Omega_i$, this is we should
explain what we mean by $\Delta v$. This function is defined by
the equality $\Delta v:=\Delta v_i$, $x\in\Omega_i$. It is
obvious that the function $\Delta v$ defined in such way is an
element of $L_2(\Pi_a)$. The set $\Pi_a$ possesses a cone
property, this is why an embedding $W_2^1(\Pi_a)\subset
L_2(\Pi_a)$ is compact due to Rellich-Kondrashov theorem
\cite[Theorem 6.2, Ch. VI]{Ad}. The right hand side in the
boundary condition (\ref{3.5}) is a trace of a function
belonging to $W_2^1(\Pi_a)$. This function can be chosen as
$v(x)\chi_{2}(x_1)$, where $\chi_{2}(x_1)$ is an infinitely
differentiable cut-off function being equal to one as
$|x_1|>(2a+l)/3$ and vanishing as $|x_1|<(a+2l)/3$. Due to
Lemma~\ref{lm3.6} the functions $v_i$ being elements of the
spaces $W_2^2(\Omega_i^a)$, it yields that $v\chi_3\in
W_2^1(\Pi_a,\Gamma^a_l)$. The equality $v(x)\chi_3(x_1)=v(x)$,
$x\in\partial\Pi_a$ follows from the definition of the function
$\chi_3$ and the relation $v(x)=0$, $x\in\Gamma_l$. Employing
the aforementioned facts, and following the idea of the proof of
Theorem~10 in \cite[Ch. I\!V,\S 1.8]{Ld}, one can check easily
that the boundary value problem (\ref{3.5}) is uniquely solvable
in the space $W_2^1(\Pi_a,\Gamma_l^a)$.

The problem (\ref{3.5}) is uniquely solvable in the space
$W_2^1(\Pi_a,\Gamma_l^a)$ (see, for instance, \cite{Ld}).
Moreover, $w$ is an element of the space $W_2^2(Q)$ for each
$Q\in\Xi$ due to theorems on improving the smoothness of
solutions to elliptic boundary value problems. In particular, it
means that in addition to the integral equality corresponding to
the problem (\ref{3.5}) (see Remark~\ref{rm1.1}) the equation in
(\ref{3.5}) holds also as the equality of two functions from
$L_2(Q)$ for each $Q\in\Xi$. Therefore, $\Delta w\in L_2(\Pi_a)$
and the equation in (\ref{3.5}) holds also as the equality of
two functions from $L_2(\Pi_a)$. The function $w$ can be also
considered as a value of linear bounded operator $T_5:
\bigoplus\limits_{i=1}^4
W_2^2(\Omega_i^a,\partial\Omega_i^a\cap\partial\Omega_i)\to
W_2^1(\Pi_a,\Gamma_l^a)$, $T_5 v:=w$. It is clear that $T_5$ is
also a linear bounded operator from $\bigoplus\limits_{i=1}^4
W_2^2(\Omega_i^a,\partial\Omega_i^a\cap\partial\Omega_i)$ into
$W_2^2(Q)$ for each $Q\in\Xi$.

Let $\chi_3(x_1)$ be  an infinitely differentiable cut-off even
function which equals minus one as $|x_1|<(a+2l)/3$ and vanishes
as $|x_1|>(2a+l)/3$. We construct the function $u$ by the rule:
\begin{equation}\label{3.6}
u(x,k):=w(x,k)\chi_3(x_1)+v(x,k)(1-\chi_3(x_1)).
\end{equation}
The function $u$ can also be regarded as $u=T_6(k)g$ where
$T_6(k)$ is a linear bounded operator from $L_2(\Pi_a)$ into
$W_2^1(\Pi_b,\Gamma_l^b)$ and $W_2^2(Q)$ for any $b>0$ and each
$Q\in\Xi$. Moreover, $T_6(\cdot)\in
\mathcal{H}(L_2(\Pi_a),W_2^1(\Pi_b,\Gamma_l^b))$ and
$T_6(\cdot)\in \mathcal{H}(L_2(\Pi_a),W_2^2(Q))$.

Let us apply the operator $-(\Delta+1-k^2)$ to $u$ and take into
account the equations for $v$ and $w$ (see (\ref{3.3}),
(\ref{3.5})). As a result we get:
\begin{equation}\label{3.7}
-(\Delta+1-k^2)u=g+(v-w)(\Delta+1-k^2)\chi_3+
2\left(\nabla\chi_3,\nabla
(v-w)\right)_{\mathbb{R}^2}=g+T_7(k)g.
\end{equation}
The function $u$ defined by (\ref{3.6}) satisfies the
homogeneous Dirichlet condition on $\partial\Pi$ and asymptotics
formulas (\ref{3.2}). Therefore, this function is a solution to
the boundary value problem (\ref{3.1}), (\ref{3.2}) if and only
if it meets the equation from (\ref{3.1}). Due to (\ref{3.7})
this leads us to the equation for the function $g$:
\begin{equation}\label{3.8}
g+T_7(k)g=f.
\end{equation}
Completely by analogy with Propositions~3.1~and~3.2 from
\cite{BEG} one can prove the following lemma.
\begin{lemma}\label{lm3.1}
The operator $T_7(k)$ is a linear compact operator from
$L_2(\Pi_a)$ into $L_2(\Pi_a)$ for each $k\in{B}$ and
$T_7(\cdot)\in\mathcal{H}(L_2(\Pi_a),L_2(\Pi_a))$. For each
$k\in{B}$ the equation (\ref{3.8}) is equivalent to the boundary
value problem (\ref{3.1}), (\ref{3.2}). Namely, for each
solution $g$ of the equation (\ref{3.8}) there exists a solution
to the boundary value problem (\ref{3.1}), (\ref{3.2}) given by
the formula $u=T_6(k)g$. For each solution $u$ to the boundary
value problem (\ref{3.1}), (\ref{3.2}) there exists a unique
solution of the equation (\ref{3.8}) associated with $u$ by the
equality $u=T_6(k)g$.
\end{lemma}

The operator $T_7$ being compact, Fredholm alternatives can be
applied to the equation (\ref{3.8}). Due to Lemma~\ref{lm3.1}
this solves the solvability questions for the boundary value
problem (\ref{3.1}), (\ref{3.2}). It should be also noted that
in the case of unique solvability of the equation (\ref{3.8})
the solution $u$ to the problem (\ref{3.1}), (\ref{3.2})
generated by the rule $u=T_6(k)(I+T_7(k))^{-1}f$ from the
solution of the equation (\ref{3.8}), coincides with the
function $({H}_l-1+k^2)^{-1}f$ as $\mathrm{Re}\, k>0$ (see the
asymptotics formulas (\ref{3.2})). This is why the operator
$T_6(k)(I+T_7(k))^{-1}$ can be interpreted as an analytic
continuation of the operator $({H}_l-1+k^2)^{-1}f$. At the same
time it should be stressed that as $\mathrm{Re}\, k\leqslant0$
the function $u=T_6(k)(I+T_7(k))^{-1}f$, generally speaking, is
not an element of the space $L_2(\Pi)$.

\begin{lemma}\label{lm3.2}
There exists a point $k_*\in{B}$ such that the operator
$(I+T_7(k_*))$ has a bounded inverse.
\end{lemma}
\begin{proof}
It is clear that it is sufficient to find the point $k_*\in{B}$
for which the equation (\ref{3.8}) is uniquely solvable. The
unique solvability of the latter is equivalent to that of the
boundary value problem (\ref{3.1}), (\ref{3.2}). We choose a
point $k_*$ as $k_*=\delta(1+\mathrm{i})$, $\delta>0$. For such
$k_*$ the problem (\ref{3.1}) with $f=0$ has no nontrivial
solution meeting the asymptotics formulas (\ref{3.2}), since
otherwise this function would be an element $L_2(\Pi)$, and
$\lambda_*=1-k_*^2$ would be a complex-valued eigenvalue of the
operator ${H}_l$. This contradicts to the reality of the
spectrum of the operator ${H}_l$.
\end{proof}

The proven lemma, compactness and holomorphy of the operator
$T_7(k)$ allow us to apply Theorem~7.1 from \cite[Ch. 15, \S
7]{SP} to the operator $(I+T_7(k))^{-1}$, what leads us to the
following statement.

\begin{lemma}\label{lm3.3}
The belonging $(I+T_7(\cdot))^{-1}\in
\mathcal{M}(L_2(\Pi_a),L_2(\Pi_a))$ takes place.
\end{lemma}

Due to this lemma the only possible singularities of the
operator $(I+T_7(k))^{-1}$ are isolated poles. We are
interesting only on presence and absence of the pole at the
point $k=0$. This is why we suppose that the neighbourhood ${B}$
of zero contains no poles except possible pole at zero. The
presence of pole at zero implies the existence of a nontrivial
solution of the equation (\ref{3.8}) with $k=0$, $f=0$, what is
equivalent to the existence of the bounded nontrivial solution
of the problem (\ref{3.1}) (see asymptotics (\ref{3.2})) with
$k=0$, $f=0$. The next lemma describes possible options  of such
solutions to exist.
\begin{lemma}\label{lm3.4}
Let $k=0$, $f=0$. Then
\begin{enumerate}
\item \label{it1lm3.4}
The boundary value problem (\ref{3.1}) has at most one
nontrivial solution meeting the asymptotics formulas (\ref{3.2})
and being even on $x_2$ in the case $d=\pi$. This solution has a
definite parity on $x_1$.
\item  \label{it2lm3.4}
If $d=\pi$, then the boundary value problem (\ref{3.1}) has a
unique nontrivial solution which is odd on $x_2$ and meets the
asymptotics formulas (\ref{3.2}), where $c_+(0)=1$. This
solution is $\sin x_2$.
\end{enumerate}
\end{lemma}
\begin{proof}
As $k=0$ the boundary value problem (\ref{3.1}) being equivalent
to the equation (\ref{3.8}), owing to compactness of the
operator $T_7(0)$ the problem (\ref{3.1}) can have only finitely
many bounded linear independent solutions. Boundedness in this
case is an implication of the asymptotics (\ref{3.2}). We denote
these solutions by $u_j$, $j=1,\ldots,q$. The change of
variables $x_1\mapsto -x_1$ maps a solution to the problem
(\ref{3.1}) into a solution, this is why without loss of
generality we can assume that each of solutions $u_j$ has a
definite parity on $x_1$. In the case $d=\pi$ we also assume
that each of these solutions is even on $x_2$. Moreover, all the
functions $u_j$ can be supposed to be real. We also observe that
due to theorems on improving smoothness we have $u_j\in
C^\infty(\overline{Q})$ for each $Q\in\Xi$.

First we prove that the coefficients $c_\pm(0)$ are always
non-zero. Suppose the opposite, namely, let there exists a
nontrivial solution $u=u_j $ whose coefficients  $c_\pm(0)$ are
zero. In view of asymptotics (\ref{3.2}) it means that the
function $u$ decays exponentially as $|x_1|\to\infty$,
$x_2\in(-d,\pi)$. Let the function $u$ be even on $x_1$. We
introduce the function
\begin{equation}\label{3.9}
U(x):=x_1\int\limits_{0}^{x_1}u(t,x_2)\,dt.
\end{equation}
The function $U$ is surely to be infinitely differentiable at
all interior points of $\Pi$. Moreover, it is an element of the
space $W_2^1(\Pi_a,\Gamma_l^a)$ for any $a>0$. Since the
function  $u$ is even on $x_1$, it follows that its derivative
on $x_1$ vanishes as $x_1=0$. Taking into account this fact and
the equation for $u$, it is not difficult to check that the
function $U$ is a solution to the equation
\begin{equation*}
(\Delta+1)U=2u,\quad x\in\Pi.
\end{equation*}
Moreover, the function $U$ satisfies the homogeneous Dirichlet
condition on the lines $x_2=-d$ and $x_2=\pi$. We are going to
prove that it vanishes on $\Gamma_l$ as well. In order to do it,
due to evenness of $u$ on $x_1$, it is sufficient to establish
the equality:
\begin{equation*}
\int\limits_{\gamma_l} u\,dx_1=0.
\end{equation*}
This equality can be proved easily by integration by parts:
\begin{equation}\label{3.12}
0=\int\limits_{\Pi^+} \sin
x_2(\Delta+1)u\,dx=\int\limits_{\gamma_l}u\,dx_1.
\end{equation}
Here we have also used the boundary condition for the function
$u$ and its exponential decaying at infinity. The function $U$
behaves like $\mathcal{O}(x_1)$ as $x_1\to\pm\infty$, what
follows from the exponential decaying of  $u$ at infinity.

Bearing in mind the properties of the functions $u$ and $U$, we
can integrate by parts:
\begin{equation*}
0=\int\limits_{\Pi}U(\Delta+1)u\,dx=
2\int\limits_{\Pi}|u|^2\,dx,
\end{equation*}
what implies $u=0$. The same equality can be also proved in the
case the function $u$ being odd on $x_1$. Here the function $U$
should be defined as
\begin{equation*}
U(x):=\int\limits_{0}^{x_1}t u(t,x_2)\,dt.
\end{equation*}
This function possesses the same properties as the function $U$
in (\ref{3.9}). The only difference in the proof of these
properties is a modification of (\ref{3.12}), in this case the
integral $0=\int\limits_{\Pi^+} x_1\sin x_2(\Delta+1)u\,dx$
should be taken as a source integral for integration by parts.
It should be also noted that the function $U$ in this case is
bounded as $x_1\to+\infty$.

Thus, each of the functions $u_j$ has nonzero coefficients
$c_\pm(0)$ in the asymptotic formulas (\ref{3.2}). It means that
the number of the functions $u_j$ does not exceed two. Indeed,
otherwise it would be possible to change a linear combination of
the functions $u_j$ whose coefficients $c_\pm(0)$ would be zero.
It would mean that this combination is identically zero, and, as
a result, that the functions $u_j$ are linear dependent. It also
obvious that in the case two functions $u_j$ are present, they
have different parity on $x_1$. Let us stress that for $d=\pi$
the assumed parity of $u_j$ on $x_2$ is essential in these
arguments otherwise the possible number of the functions $u_j$
increases up to four.

Let the number of the functions $u_j$ be two and let $u_1$ be
even on $x_1$ and $u_2$ be odd. Without loss of generality we
assume that the coefficients  $c_\pm(0)$ of the functions $u_1$
and $u_2$ are respectively of the form $c_\pm(0)=\sqrt{2/\pi}$,
$c_\pm(0)=\pm \sqrt{2/\pi}$. We set
\begin{equation*}
U_1(x):=\int\limits_0^{x_1} u_1(t,x_2)\,dt.
\end{equation*}
By analogy with how the properties of the function $U$ in
(\ref{3.9}) have been found, one can show easily that $U_1$ is a
solution to the boundary value problem (\ref{3.1}) meets the
asymptotics representation
\begin{equation*}
U_1(x)=\sqrt{\frac{2}{\pi}}(x_1\pm c)\sin
x_2+\mathcal{O}(\mathrm{e}^{-\sqrt{3}|x_1|}),\quad
x_1\to\pm\infty,\quad x_2\in(0,\pi),
\end{equation*}
where $c$ is some constant. In the case $d=\pi$ the function
$U_1$ has exactly the same asymptotics as $x_2\in(-\pi,0)$. If
$d<\pi$, then the function $U_1$ decays exponentially as
$x_1\to\pm\infty$, $x_2\in(-d,0)$, what follows from the
boundary value problem for $U_1$ and boundedness of $U_1$ as
$x_1\to\pm\infty$, $x_2\in(-d,0)$. Taking into account the
properties of the functions $U_1$ and $u_2$, integrating by
parts in an integral $\int\limits_{\Pi_R} U_1(\Delta+1)u_2\,dx$
and passing after that to limit as $R\to+\infty$, we get:
\begin{equation*}
0=\int\limits_{\Pi} U_1(\Delta+1)u_2\,dx=\left\{
\begin{aligned}
&-2,&&d<\pi,
\\
&-4,&&d=\pi,
\end{aligned} \right.
\end{equation*}
a contradiction. Thus, the number of the functions $u_j$ is at
most one and if exists, this function is unique and has a
definite parity on $x_1$. Item~\ref{it1lm3.4} is proven.

Statement of Item~\ref{it2lm3.4} is obvious if one takes into
account that odd on $x_2$ solution vanishes as $x_2=0$.
\end{proof}

Let us introduce auxiliary notations. In the case the nontrivial
solution to the problem (\ref{3.1}) described in
Item~\ref{it1lm3.4} of Lemma~\ref{lm3.4} exists, we denote this
solution by $\phi(x)$. The associated solution of the equation
(\ref{3.8}) is indicated as $\Phi(x)$,
$\phi(x)=(T_6(0)\Phi)(x)$. If such solution does not exists, we
set $\phi=0$, $\Phi=0$. In the case $d=\pi$ the solution of the
equation (\ref{3.8}) associated with $\sin x_2$ is denoted by
$\widetilde\Phi(x)$, $\sin x_2=(T_6(0)\widetilde\Phi)(x)$.

The next lemma describes the structure of the operator
$(I+T_7(k))^{-1}$ for small $k$.

\begin{lemma}\label{lm3.5}
The operator $(I+T_7(k))^{-1}$ can be represented as:
\begin{align*}
&(I+T_7(k))^{-1}=\frac{1}{k}T_8+T_9(k),
\\
&T_8f:=\frac{1}{2}\Phi\int\limits_{\Pi}f(x)\phi(x)\,dx,&&\text{as
$d<\pi$},
\\
&T_8f:=\frac{1}{4}\Phi\int\limits_{\Pi}f(x)\phi(x)\,dx+
\frac{1}{2\pi} \widetilde\Phi\int\limits_\Pi {f(x)\sin x_2}\,dx,
&&\text{as $d=\pi$},
\end{align*}
where $T_9(\cdot)\in
\mathcal{H}\left(L_2(\Pi_a),L_2(\Pi_a)\right)$.
\end{lemma}

The proof of this lemma is analogous to that of Theorem~3.4 in
\cite{BEG}.

\section{Asymptotics expansions of emerging eigenvalues}

In the present section we will prove
Items~\ref{th1.1.it3},~\ref{th1.1.it4} of Theorem~\ref{th1.1}
finishing by this the proof of this theorem. For calculating the
asymptotics expansions we will employ the scheme which is
analogous to that employed in \cite{BEG} in the case $d=\pi$.
The main ideas of this scheme are borrowed from the works
\cite{G1}, \cite{G2}. Let $l_*$ be some value of the length of
the window $\gamma_l$. We give an increment
$\varepsilon\in(-\varepsilon_0,\varepsilon_0)$ to this length.
Here $\varepsilon_0$ is from (\ref{2.0}). As it was shown in the
proof of Lemma~\ref{lm2.1}, the eigenvalues of the operator
$H_{l_*+\varepsilon}$ are those of the boundary value problem
(\ref{2.1}). We denote $\lambda=1-k^2$, then in accordance with
the results of the previous section the boundary value problem
(\ref{2.1}) is equivalent to an operator equation in
$L_2(\Pi_a)$:
\begin{equation}\label{4.1}
(I+T_7(k)-\varepsilon L_\varepsilon T_6(k))g=0.
\end{equation}
Here the parameter $a$ should be chosen great enough and
independent on $\varepsilon$ so that the supports of the
coefficients of the operator $L_\varepsilon$ lie inside $\Pi_a$
for all $\varepsilon$ small enough.

Since $T_6(\cdot)\in \mathcal{H}(L_2(\Pi_a),W_2^2(Q))$ for each
$Q\in\Xi$, in view of the form of the coefficients of the
operator $L_\varepsilon$ (see (\ref{2.1})) we conclude that
$L_\varepsilon T_6(\cdot)\in
\mathcal{H}(L_2(\Pi_a),L_2(\Pi_a))$. Moreover, the operator
$L_\varepsilon T_6(k)$ is bounded uniformly on $\varepsilon$.

The question on existence of the eigenvalues of the operator
${H}_l$ emerging from the threshold of the essential spectrum is
surely to be equivalent to the question on existence of the
function $k=k_\varepsilon\xrightarrow[\varepsilon\to0]{}0$ so
that the equation (\ref{4.1}) have a nontrivial solution
$g_\varepsilon$ such that $T_6(k_\varepsilon)g_\varepsilon\in
L_2(\Pi)$. This is why it is sufficient to study the question on
existence of such function $k_\varepsilon$.

\begin{lemma}\label{lm4.1}
Let for $l=l_*$ there exist no solution $\phi$ described in
Item~\ref{th1.1.it3} of Theorem~\ref{th1.1}. Then there exist
$\varepsilon_0>0$ and $\delta>0$ such that as
$|l-l_*|<\varepsilon_0$ the operator $H_l$ has no eigenvalues in
an interval $(1-\delta,1+\delta)$.
\end{lemma}
\begin{proof}
We begin with the case $d<\pi$. In accordance with this
assumption and Lemma~\ref{lm3.5} the operator $(I+T_7(k))$ is
invertible for each $k\in {B}$. Since the operator
$L_\varepsilon T_6(k)$ is bounded uniformly on $\varepsilon$ and
$k\in {B}$,  for sufficiently small $\varepsilon$ the operator
in (\ref{4.1}) is also invertible for each $k\in{B}$. Therefore,
the equation (\ref{4.1}) has no nontrivial solutions.

In the case $d=\pi$ the proof is analogous. The set $\Pi$ being
symmetric w.r.t. the axis $x_2=0$, all the eigenfunctions of the
operator $H_{l}$ are even on $x_2$, since odd eigenfunctions
would satisfy Dirichlet condition on $x_2=0$ and would be the
eigenfunctions of the operator $H_0$. At the same time, the
discrete spectrum of the latter is empty. Taking into account
the parity of the eigenfunctions on $x_2$, it is sufficient to
consider the equation (\ref{4.1}) on even on $x_2$ functions $g$
only (clear, the operator $T_6(k)$ preserves the parity on
$x_2$). We denote by $\mathcal{V}$ the subspace of $L_2(\Pi_a)$
consisting of even on $x_2$ functions. Then the operator
$(I+T_7(k))^{-1}L_\varepsilon T_6(k)$ $: \mathcal{V}\to
L_2(\Pi_a)$ is bounded uniformly on $\varepsilon$ (see
Lemma~\ref{lm3.5}). Using this fact, one can easily deduce the
absence of nontrivial even on $x_2$ solution of the equation
(\ref{4.1}).
\end{proof}

Now we are going to prove that the existence of nontrivial
solution $\phi$ from Lemma~\ref{lm3.5} for $l=l_*$ implies the
existence of the function
$k=k_\varepsilon\xrightarrow[\varepsilon\to0]{}0$ so that the
equation (\ref{4.1}) have a nontrivial solution. We are also
going to show that the function $k_\varepsilon$ meets the
equality
\begin{equation}\label{4.2}
k_\varepsilon=\varepsilon\mu+\mathcal{O}(\varepsilon^2),
\end{equation}
where $\mu$ is defined by the formulas (\ref{1.4}), (\ref{1.4a})
with $l_n$ and $\phi_n$ replaced by $l_*$, $\phi$, respectively.
We adduce the proof in the case $d<\pi$ only; the case $d=\pi$
was proven in \cite{BEG}.

In the equation (\ref{4.1}) we invert the operator $(I+T_7(k))$
taking into account Lemma~\ref{lm3.5}:
\begin{equation*}
g-\frac{\varepsilon}{2k}\Phi\int\limits_{\Pi}\phi L_\varepsilon
T_6(k)g\,dy+\varepsilon T_9(k)L_\varepsilon T_6(k)g=0.
\end{equation*}
The operator $T_9(k)L_\varepsilon T_6(k)$ is bounded uniformly
on $\varepsilon$, this is why for sufficiently small
$\varepsilon$ there exists a bounded inverse
$T_{10}(k,\varepsilon):=(I+\varepsilon T_9(k) L_\varepsilon
T_6(k))^{-1}$. Applying this operator to the latter equation, we
obtain:
\begin{equation}\label{4.4}
g-\frac{\varepsilon}{2k}\left(\int\limits_\Pi\phi L_\varepsilon
T_6(k)g\,dy\right) T_{10}(k,\varepsilon)\Phi=0.
\end{equation}
If the integral in the right-hand side is zero, it immediately
leads us to the trivial solution $g=0$. Therefore, on a
nontrivial solution this integral is nonzero. Bearing in mind
this fact, we apply the operator $L_\varepsilon T_6(k)$ to the
equation (\ref{4.4}), multiply then by $2k\phi$ and integrate
over $\Pi$. This results in the following equation:
\begin{equation}\label{4.5}
2k-\varepsilon\int\limits_\Pi \phi L_\varepsilon T_6(k)
T_{10}(k,\varepsilon)\Phi\,dy=0.
\end{equation}
In fact, this is an equation for the function $k=k_\varepsilon$.
Due to (\ref{4.4}) the corresponding nontrivial solution of the
equation (\ref{4.1}) is given by the formula:
\begin{equation}\label{4.6}
g_\varepsilon=C T_{10}(k_\varepsilon,\varepsilon)\Phi,
\end{equation}
where $C$ is an arbitrary constant. The function
\begin{equation*}
(k,\varepsilon)\mapsto \varepsilon\int\limits_\Pi \phi
L_\varepsilon T_6(k) T_{11}(k,\varepsilon)\Phi\,dy
\end{equation*}
is holomorphic on $k$ and tends to zero as $\varepsilon\to0$
uniformly on $k$. Therefore, on the boundary of the domain ${B}$
it will be less by absolute value than  $2|k|$ if $\varepsilon$
is small enough. By Rouche theorem it follows that for
sufficiently small $\varepsilon$ the equation (\ref{4.5}) has
the same number of roots in ${B}$ as the number of zeros for the
function $k \mapsto 2k$, i.e., the unique root. We denote this
root by $k_\varepsilon$. Clear, the convergence
$T_{10}(k,\varepsilon)\xrightarrow[\varepsilon\to0]{} I$ holds
true in the operator norm uniformly on $k\in {B}$. In view of
the equality $\phi=T_6(0)\Phi$ and (\ref{2.1}) it allows to the
rewrite the equation (\ref{4.5}) as
\begin{align*}
&k_\varepsilon=\frac{\varepsilon}{2}\int\limits_\Pi\phi
L_0\phi\,dy+\mathcal{O}(\varepsilon
|k_\varepsilon|+\varepsilon^2),
\\
&L_0:=-2\chi'_1(y_1)\frac{\partial^2}{\partial
y_1^2}-\chi''_1(y_1)\frac{\partial}{\partial y_1}.
\end{align*}
Here we have also taken into account the form of the
coefficients of the operator $L_\varepsilon$. Since
$k_\varepsilon\to0$, the equalities obtained imply that
$k_\varepsilon=\mathcal{O}(\varepsilon)$. Hence,
\begin{equation}\label{4.6a}
k_\varepsilon=\frac{\varepsilon}{2}\int\limits_\Pi\phi
L_0\phi\,dy+\mathcal{O}(\varepsilon^2).
\end{equation}

\begin{lemma}\label{lm4.2}
In a vicinity of the right edge of the window $\gamma_l$ the
function $\phi(x)$ behaves as:
\begin{equation}\label{4.15}
\phi(y)=\alpha r^{1/2}\sin\frac{\theta}{2}+\mathcal{O}(r),\quad
\frac{\partial}{\partial y_i}\phi(y)=\alpha
\frac{\partial}{\partial
y_i}r^{1/2}\sin\frac{\theta}{2}+\mathcal{O}(1),\quad r\to0,
\end{equation}
where $(r,\theta)$ are polar coordinates centered at the right
edge of the window $\gamma_l$, $\alpha$ is a some constant.
\end{lemma}

\begin{proof}
Let $\chi_4=\chi_4(r)$ be an infinitely differentiable cut-off
function which equals one as $r\leqslant\delta$ and vanishes as
$r\geqslant 2\delta$. We denote $\Theta:=\{y: r<2\delta,
0<\theta<2\pi\}$. We choose the number $\delta$ so that the
circle $\overline{\Theta}$ not to intersect the left edge of the
window $\gamma_l$ and lie inside $\Pi$. As it was mentioned in
Remark~\ref{rm1.1}, $\phi\in C^\infty(\Pi)$. Taking into account
this fact, one can easily check that the function
$\widetilde\phi(y)=\chi_4(r)\phi(y)\in
W_2^1(\Theta,\partial\Theta)$ is a solution to the boundary
value problem:
\begin{equation*}
-\Delta\widetilde\phi=\lambda\widetilde\phi+\widetilde f,\quad
y\in\Theta, \qquad \widetilde\phi=0,\quad y\in\partial\Theta,
\end{equation*}
where $\widetilde f\in C^\infty(\overline{\Theta})$,
$\widetilde{f}\equiv0$ as $r\leqslant \delta$. Changing
variables in this problem $\widetilde r=r^{1/2}$,
$\widetilde\theta=\theta/2$, we arrive at the following boundary
value problem:
\begin{equation*}
-\Delta_{\widetilde y}\widetilde\phi=4\lambda{\widetilde
r}^2\widetilde\phi+4{\widetilde r}^2\widetilde f,\quad
\widetilde y\in\widetilde\Theta, \qquad \widetilde\phi=0,\quad
\widetilde y\in\partial\widetilde\Theta,
\end{equation*}
where $\widetilde y$ are Cartesian coordinates associated with
$(\widetilde r,\widetilde\theta)$,
$\widetilde\Theta:=\{\widetilde y: \widetilde r<\sqrt{2\delta},
\widetilde\theta\in(0,\pi)\}$. Since $\widetilde\phi\in
W_2^1(\Theta )$,
\begin{equation*}
\int\limits_{\widetilde\Theta}|\nabla_{\widetilde
y}\widetilde\phi|^2\,\mathrm{d} \widetilde
y=\int\limits_{\Theta}|\nabla_{y}\widetilde\phi|^2\,\mathrm{d}
y<\infty
\end{equation*}
and the function $\widetilde\phi$ vanishes at the boundary of
the domain $\widetilde\Theta$, the belonging $\widetilde\phi\in
W_2^1(\widetilde\Theta,\partial\widetilde\Theta)$ holds true. We
continue the functions $\widetilde{f}$ and $\widetilde{\phi}$
into the domain $\{y:
\widetilde{r}<2\delta,\pi<\widetilde{\theta}<2\pi\}$ as follows:
$\widetilde{\phi}(\widetilde{y}_1,\widetilde{y}_2)=
-\widetilde{\phi}(-\widetilde{y}_1,\widetilde{y}_2)$,
$\widetilde{f}(\widetilde{y}_1,\widetilde{y}_2)=
-\widetilde{f}(-\widetilde{y}_1,\widetilde{y}_2)$,
$\widetilde{y}_2<0$. We preserve former notations
$\widetilde{\phi}$ and $\widetilde{f}$. for the functions
continued. We denote $\widehat{\Omega}:=\{\widetilde{y}:
r<\sqrt{2\delta}\}$. It is clear that $\widetilde{f}\in
L_2(\widehat{\Omega})$, $\widetilde{f}\equiv0$ as
$r\leqslant\sqrt{\delta}$, and the function $\widetilde{\phi}\in
W_2^1(\widehat{\Omega},\partial\widehat{\Omega})$ is a solution
to the boundary value problem:
\begin{equation*}
-\Delta\widetilde{\phi}=4r^2\widetilde{\phi}+\widetilde{f},\quad
\widetilde{y}\in\widehat{\Omega},\qquad \widetilde{\phi}=0,\quad
\widetilde{y}\in\partial\widehat{\Omega}.
\end{equation*}
Due to the theorems on improving smoothness of solutions to
elliptic boundary value problems (see \cite[Ch. 4, \S 2]{Ld})
the function $\widetilde\phi$ is infinitely differentiable at
zero. Moreover, in view of the boundary value problem for
$\widetilde\phi$ we have:
\begin{equation*}
\widetilde\phi(\widetilde y)=\alpha\widetilde
y_2+\mathcal{O}(\widetilde r^2),\quad
\frac{\partial}{\partial\widetilde
y_i}\widetilde\phi(y)=\alpha\frac{\partial\widetilde
y_2}{\partial\widetilde y_i}+\mathcal{O}(\widetilde r),\quad
\widetilde r\to0.
\end{equation*}
Returning now to the variables $y$ and taking into account the
definition of the function $\chi_4$, we arrive at the statement
of the lemma.
\end{proof}

\begin{remark}\label{rm4.1}
The idea of the proof of Lemma~\ref{lm4.2} is borrowed from the
proof of Lemma~3.1 in \cite[Ch. I\!I\!I, \S 2]{Il}.
\end{remark}

The function $\phi$ has a definite parity on $x_1$, this is why
the formulas similar to (\ref{4.15}) are also valid for the left
edge of the window $\gamma_l$. Taking into account the behaviour
of the function $\phi(x)$ in the vicinities of the edges of the
window $\gamma_l$ and the boundary value problem for $\phi$, we
can evaluate the integral in (\ref{4.6a}) by integrating by
parts twice:
\begin{equation}\label{4.9}
\int\limits_\Pi\phi
L_0\phi\,dy=-\int\limits_\Pi\phi\left(\Delta+1\right)
\left(\chi(y_1)\frac{\partial\phi}{\partial
y_1}\right)\,dy=\pi\alpha^2.
\end{equation}
In the same way we check that
\begin{equation}\label{4.8a}
0=\int\limits_\Pi y_1\frac{\partial\phi}{\partial
y_1}\left(\Delta+1\right)\phi \,dy=\pi
l_*\alpha^2+2\int\limits_{\Pi}\phi\frac{\partial^2\phi}{\partial
y_1}\,dy=\pi
l_*\alpha^2-2\int\limits_\Pi\left|\frac{\partial\phi}{\partial
y_1 }\right|^2\,dy,
\end{equation}
what together with (\ref{4.9}) imply the equality:
\begin{equation}\label{4.9a}
\int\limits_\Pi\phi L_0\phi\,dy=\frac{2}{l_*}
\int\limits_{\Pi}\phi\frac{\partial^2\phi}{\partial y_1}\,dy
=2\mu,
\end{equation}
where $\mu$ is defined by the formula (\ref{1.4}) with $l_n$ and
$\phi_n$ replaced by $l_*$ and $\phi$. Substituting the
relations (\ref{4.9a}) into (\ref{4.6a}), we arrive at the
asymptotics (\ref{4.2}).

We put $C=1$ in (\ref{4.6}), then in view of the form of the
operator $T_{10}(k,\varepsilon)$ the obtained solution of the
equation (\ref{4.1}) satisfies an asymptotic formula:
\begin{equation}\label{4.10}
g_\varepsilon=\Phi+\mathcal{O}(\varepsilon)\quad \text{in the
norm of $L_2(\Pi_a)$}.
\end{equation}
Due to Lemma~\ref{lm3.1} and relation $\Phi\not\equiv0$ this
equality implies that the function
$\psi^\varepsilon(y)=(T_6(k_\varepsilon)g_\varepsilon)(y)$ is
not identically zero. Therefore, it is an eigenfunction of the
boundary value problem (\ref{2.1}). Due to the asymptotics
representations (\ref{3.2}), (\ref{4.2}) and (\ref{4.9a}) the
inequality $\mathrm{Re}\,k_\varepsilon>0$ holds true only as
$\varepsilon>0$. Hence, the function $\psi^\varepsilon$ is an
element of $L_2(\Pi)$ only as $\varepsilon>0$. Passing to the
variables $x$ (see (\ref{2.0})), we conclude that a quantity
$\lambda^\varepsilon:=1-k_\varepsilon^2$  is an eigenvalue of
the operator ${H}_{l_*+\varepsilon}$ only as $\varepsilon>0$,
and $\psi^\varepsilon(y(x,\varepsilon))$ is the associated
eigenfunction in this case. As $\varepsilon\leqslant 0$ the
operator $H_{l_*+\varepsilon}$ has no eigenvalues close to the
threshold of the essential spectrum, i.e., the eigenvalue
$\lambda^\varepsilon$ disappears as $\varepsilon\leqslant0$.
Therefore, $l_*$ is a critical value of the length of the window
$\gamma_l$, and the corresponding eigenvalue emerging as $l>l_*$
has the asymptotics (\ref{1.4b}), (\ref{1.4}), what follows from
(\ref{4.2}) and (\ref{4.9a}). We assume that $l_*=l_n$, then
$\lambda_n=\lambda^\varepsilon$,
$\psi_n(x)=c_\varepsilon\psi^\varepsilon(y(x,\varepsilon))$,
where $c_\varepsilon$ is a some constant.

To finish the proof we need just to establish the relationships
(\ref{1.5}), (\ref{1.6}). The functions $\psi_n$ and $\phi_n$
having the same parity on $x_1$ follows from (\ref{1.6}) and
Item~\ref{th1.1.it1} of Theorem~\ref{th1.1}.

The function $\psi_n(x)$ introduced above meets the asymptotics
formulas (\ref{3.2}), where
$c_+(k_\varepsilon)=c_\varepsilon\left(\sqrt{2/\pi}
+\mathcal{O}(\varepsilon)\right)$. This equality is implied by
(\ref{4.10}), the definition of the operator $T_6(k)$ and the
functions $\phi$ and $\Phi$, the asymptotics for $k_\varepsilon$
established above, and the equality $y_1(x_1,\varepsilon)\equiv
x_1$ as $x_1$ large enough. Thus, the constant $c_\varepsilon$
can be chosen such that the function $\psi_n$ to satisfy the
asymptotics expansion (\ref{1.5}). Moreover, in this case we
have $c_\varepsilon=1+\mathcal{O}(\varepsilon)$. This equality
and (\ref{4.10}) yield:
\begin{equation*}
\psi_n(x)=\phi_n(y(x,\varepsilon))+\mathcal{O}(\varepsilon)
\end{equation*}
in the norm of $W_2^1(\Pi_R)$ for each $R>0$ (the norm here is
treated in the sense of the variables $x$). Thus, in order to
prove the assertions (\ref{1.5}) it is sufficient to check that
\begin{equation}\label{4.11}
\|\phi_n(x)-\phi_n(y(x,\varepsilon))\|_{W_2^1(\Pi_R)}=
\mathcal{O}(\varepsilon^{1/2})
\end{equation}
for each $R>0$. Clear, it is sufficient to check this equality
only as $R>l$. In the domain $\Pi_R$ we select two rectangles
$P_\pm:=\{x: \pm x_1\in(l-2\varepsilon_0,l+2\varepsilon_0),
-d<x_2<\pi\}$, where $\varepsilon_0$ is from (\ref{2.0}). We
choose $\varepsilon_0$ small enough so that the minimal
eigenvalue of Dirichlet Laplacian in the rectangles be greater
than one. We denote this eigenvalue by $\tau$. We set
$\varphi(x):=\phi_n(x)-\phi_n(y(x,\varepsilon))$. The function
$\varphi$ is an element of $W_2^1( P_\pm,\partial P_\pm)$, what
implies the estimate:
\begin{equation}\label{4.12}
\tau\|\varphi\|^2_{L_2(P_\pm)}\leqslant
\|\nabla\varphi\|^2_{L_2(P_\pm)}.
\end{equation}
Integrating by parts in the equality
\begin{equation*}
\int\limits_{P_+}\varphi(x)\left((\Delta_x+1)\varphi(x)+\varepsilon
L_\varepsilon\phi_n(y)\right)\,dx=0,
\end{equation*}
in view of properties of $\phi_n$ we obtain
\begin{equation}\label{4.13}
\begin{aligned}
\|\nabla\varphi&\|^2_{L_2(P_+)}=\|\varphi\|^2_{L_2(P_+)}+
\varepsilon\int\limits_{P_+}\varphi L_\varepsilon\phi_n(y)\,dx+
\\
&+\int\limits_{l_*-\varepsilon}^{l_*}\phi_n(x_1,0)
\left(\frac{\partial}{\partial
x_2}\phi_n(y_1(x_1,\varepsilon),-0)-\frac{\partial}{\partial
x_2}\phi_n(y_1(x_1,\varepsilon),+0)\right)\,dx_1.
\end{aligned}
\end{equation}
The latter term in the right-hand side of this equality can be
estimated taking into account (\ref{4.15}):
\begin{equation*}
\left|\int\limits_{l_*-\varepsilon}^{l_*}\phi_n(x_1,0)
\left(\frac{\partial}{\partial
x_2}\phi_n(y_1(x_1,\varepsilon),-0)-\frac{\partial}{\partial
x_2}\phi_n(y_1(x_1,\varepsilon),+0)\right)\,dx_1\right|\leqslant
C_1\varepsilon,
\end{equation*}
where $C_1$ is a some constant independent on $\varepsilon$. The
second term in the right-hand side of (\ref{4.13}) is estimated
as follows:
\begin{equation*}
\left|\int\limits_{P_+}\varphi
L_\varepsilon\psi_n(y)\,dx\right|\leqslant C_2,
\end{equation*}
where $C_2$ is a some constant independent on $\varepsilon$. In
order to establish the latter estimate one just needs to take
into account the smoothness of the function
$\phi_n(y(x,\varepsilon))$ as well as the coefficients of the
operator $L_\varepsilon$ being separated from the window
$\gamma_{l_*}$ by a positive distance uniformly on
$\varepsilon$. Substituting two last estimates into
(\ref{4.13}), we get
\begin{equation}\label{4.14}
\|\nabla\varphi\|^2_{L_2(P_+)}\leqslant
C_1\varepsilon+C_2\varepsilon\|\varphi\|_{L_2(P_+)}+
\|\varphi\|^2_{L_2(P_+)}.
\end{equation}
Similar estimate is valid for $P_-$ as well.

Bearing in mind the obtained estimates for
$\|\nabla\varphi\|^2_{L_2(P_\pm)}$, by (\ref{4.12}) we deduce:
\begin{equation*}
(\tau-1)\|\varphi\|^2_{L_2(P_\pm)}-
C_2\varepsilon\|\varphi\|_{L_2(P_\pm)}-C_1\varepsilon\leqslant
0.
\end{equation*}
Solving this square inequality with the relations $\tau>1$ and
(\ref{4.14}) taken into account, we get:
\begin{equation*}
\|\varphi\|_{L_2(P_\pm)}=\mathcal{O}(\varepsilon^{1/2}),\quad
\|\nabla\varphi\|_{L_2(P_\pm)}=\mathcal{O}(\varepsilon^{1/2}).
\end{equation*}
To finish the proof of (\ref{4.11}) it is sufficient now to note
that the estimate
\begin{equation*}
\|\varphi\|_{W_2^1\left(\Pi_R\setminus(P_+\cup
P_-)\right)}=\mathcal{O}(\varepsilon)
\end{equation*}
holds true due to the definition of the function $\varphi$ and
the function $\phi_n$ being infinitely differentiable on the set
$\overline{\Pi_R\setminus(P_+\cup P_-)}$. The proof of
Theorem~\ref{th1.1} is complete.

\section{Asymptotics expansions of the eigenvalues as
$l\to+\infty$}

This section is devoted to the proof of Theorem~\ref{th1.2}. We
begin with the proof of Item~\ref{th1.2.it1}.

In accordance with Theorem~\ref{th1.1} all the eigenfunction of
the operator $H_l$ have a certain parity on $x_1$. Thus, we may
bisect the set $\Pi$ by a segment $\{0\}\times[-d,\pi]$ and
impose on it Dirichlet or Neumann condition subject to the
parity of an eigenfunction studied. Hence, we just to need to
deal with the eigenvalue problem for the Laplacian in the right
half of $\Pi$. In this problem we make the change of the
variables by the rule $x_1\mapsto x_1-l$ what leads us to the
problem on the spectrum of the Laplacian in a domain
$\Pi^{*,l}:=\Pi\cap\{x: x_1>-l\}$ subject to appropriate
boundary conditions. Below it will be shown that as
$l\to+\infty$ such problem can be treated as a problem on
perturbation of the operator $H_*$ defined in the first section.
It will allow us to get the needed asymptotics expansions
(\ref{1.7}).

First we study the behaviour of the operator
$(H_*-\lambda)^{-1}$ as $\lambda$ close to
$\varkappa:=\frac{\pi}{\pi+d}$.  In order to do it we will
employ the same approach as that used in the third section.

We set $\Pi^*_a:=\Pi^*\cap\{x: |x_1|<a\}$,
$\lambda=\varkappa^2+k^2$. For small complex $k\in {B}$ we
consider the boundary value problem
\begin{align}
&-\Delta u=(\varkappa^2+k^2)u+f,\quad x\in\Pi^*,\qquad u=0,\quad
x\in\partial\Pi^*,\label{5.1}
\\
&
\begin{aligned}
u(x,k)&=c(k)\mathrm{e}^{\mathrm{i} k x_1}\sin
\varkappa(x_2-\pi)+\mathcal{O}(
-\mathrm{e}^{-\sqrt{3\varkappa^2-k^2}x_1}),\quad x_1\to-\infty,
\\
u(x,k)&=\mathcal{O}(\mathrm{
e}^{-\sqrt{1-\varkappa^2-k^2}x_1}),\quad\hphantom{^2.}
x_1\to+\infty,\quad x_2\in(0,\pi),
\\
u(x,k)&=\mathcal{O}(\mathrm{
e}^{-\sqrt{\frac{\pi^2}{d^2}-\varkappa^2-k^2}x_1}),\quad
x_1\to+\infty,\quad x_2\in(-d,0).
\end{aligned}\label{5.2}
\end{align}
Here $f\in L_2(\Pi^*)$ is a function whose supports lies in
$\Pi_a^*$, $a>0$ is a some fixed number, $c(k)$ is a some
constant. Let $g$ be a function from $L_2(\Pi_a^*)$ continued by
zero in $\Pi^*\setminus\overline{\Pi^*_a}$. We denote
$\Omega_0:=\Pi^*\cap\{x: x_1<0\}$. The boundary value problems
\begin{equation}\label{5.3}
-\Delta v_i=(\varkappa^2+k^2)v_i+g,\quad x\in\Omega_i, \qquad
v_i=0,\quad x\in\partial\Omega_i,\quad i=0,1,2,
\end{equation}
are solved by separation of variables:
\begin{align}
&v_i(x,k)=\sum\limits_{j=1}^\infty \int\limits_{\Omega_i}
G^i_{\!j}(x,t,k)g(t)\,dt,\label{5.4}
\\
&G_j^0(x,t,k):=\frac{1}{(\pi+d)s_j^0}\left(\mathrm{
e}^{-s_j^0|x_1-t_1|}- \mathrm{e}^{s_j^0(x_1+t_1)}\right)\sin j
\varkappa (x_2-\pi)\sin j \varkappa (t_2-\pi),\nonumber
\\
&G_j^1(x,t,k):=\frac{1}{\pi s_j^1}\left(\mathrm{
e}^{-s_j^1|x_1-t_1|}- \mathrm{e}^{-s_j^1(x_1+t_1)}\right)\sin j
x_2\sin j t_2,\nonumber
\\
&G_j^2(x,t,k):=\frac{1}{s_j^2
d}\left(\mathrm{e}^{-s_j^2|x_1-t_1|}-
\mathrm{e}^{-s_j^2(x_1+t_1)}\right)\sin \frac{\pi j}{d} x_2\sin
\frac{\pi j}{d} t_2,\nonumber
\end{align}
where $s_1^0=\mathrm{i}k$, $s_j^0=\sqrt{\varkappa^2
j^2-\varkappa^2-k^2}$, $j\geqslant 2$,
$s_j^1=\sqrt{j^2-\varkappa^2-k^2}$, $s_j^2=\sqrt{\frac{\pi^2
j^2}{d^2}-\varkappa^2-k^2}$. As $k=0$ the function $G_1^0$ is
defined by continuity:
\begin{equation*}
G_1^0(x,t,0):=-\frac{1}{(\pi+d)s_1^0}(|x_1-t_1|+x_1+t_1)\sin
\varkappa (x_2-\pi)\sin \varkappa (t_2-\pi).
\end{equation*}
We set $\Omega_0^b:=\Omega_0\cap\Pi_b$. An analogue of
Lemma~\ref{lm3.6} holds true.

\begin{lemma}\label{lm5.1}
Let $b>0$. The series (\ref{5.4}) converge in the norm of
$W_2^2(\Omega^b_i)$. The functions $v_i(x)$ meet the asymptotics
formulas (\ref{5.2}). The mapping  $g\mapsto v_i$ are linear
bounded operators from $L_2(\Pi_a^*)$ into $W_2^2(\Omega_i^b))$
as functions on $k$ belonging to
$\mathcal{H}(L_2(\Pi_a^*),W_2^2(\Omega_i^b))$.
\end{lemma}

Let $v(x,k):=v_i(x,k)$, $x\in\Omega_i$. We introduce the
function $w(x,k)$ as a solution to a boundary value problem
\begin{equation}\label{5.5}
\Delta w=\Delta v,\quad x\in\Pi_a^*, \qquad w=v,\quad
x\in\partial\Pi_a^*.
\end{equation}
Here $\Delta v$ is treated in the same sense as in (\ref{3.5}).
We denote $\Gamma^*_a:=\partial\Pi^*\cap\{x: |x_1|<a\}$. The
function $w$ can be regarded as $w=T_{11}v$, where $T_{11}:
\bigoplus\limits_{i=0}^2
W_2^2(\Omega_i^a,\partial\Omega_i^a\cap\partial\Omega_i)\to
W_2^1(\Pi_a^*,\Gamma^*_a)$ is a linear bounded operator.
Moreover, the operator $T_{11}$ is bounded as an operator from
$\bigoplus\limits_{i=0}^2
W_2^2(\Omega_i^a,\partial\Omega_i^a\cap\partial\Omega_i)$ into
$W_2^2(Q)$ for each $Q\in\Xi^*$, where $\Xi^*$ is a subset of
all bounded subdomains of  $\Pi_*$ having smooth boundary and
separated from zero by a positive distance. Let $\chi_5(x_1)$ be
an infinitely differentiable cut-off function which equals one
as $|x_1|<a/3$ and vanishes as $|x_1|>2a/3$. We define the
function $u(x,k)$ by a rule:
\begin{equation}\label{5.6}
u(x,k):=w(x,k)\chi_5(x_1)+v(x,k)(1-\chi_5(x_1)).
\end{equation}
The function $u$ is treated as a value of a linear operator
$T_{12}(k)g$ defined by a rule $T_{12}(k)g:=u$. The operator
$T_{12}: L_2(\Pi_a^*)\to W_2^1(\Pi_b^*,\Gamma_b^*)$, $T_{12}:
L_2(\Pi_a^*)\to W_2^2(Q)$ is bounded for all $b>0$ and each
$Q\in\Xi^*$. Moreover, $T_{12}(\cdot)\in
\mathcal{H}\left(L_2(\Pi_a^*), W_2^1(\Pi_b^*,\Gamma_b^*)\right)$
and $T_{12}(\cdot)\in \mathcal{H}(L_2(\Pi_a^*),W_2^2(Q))$ for
all $b>0$ and each $Q\in\Xi^*$.

By analogy with the deriving the equation (\ref{3.8}) it can be
shown that the function $u$ from (\ref{5.6}) is a solution to
the boundary value problem (\ref{5.1}), (\ref{5.2}), if $u$ is a
solution to the equation
\begin{equation}\label{5.7}
g+T_{13}(k)g=f,
\end{equation}
where
\begin{equation*}
T_{13}(k)g:=(v-w)\left(\Delta+\varkappa^2+k^2\right)\chi_5(x_1)+
2\left(\nabla_x \chi_5, \nabla_x(v-w)\right)_{\mathbb{R}^2}.
\end{equation*}
By analogy with Lemmas~\ref{lm3.1}-\ref{lm3.3} one can establish
the following statement.

\begin{lemma}\label{lm5.2}
The operator $T_{13}(k)$ is a linear compact operator from
$L_2(\Pi^*_a)$ into $L_2(\Pi^*_a)$ for all $k\in{B}$ and
$T_{13}(\cdot)\in\mathcal{H}(L_2(\Pi^*_a),L_2(\Pi^*_a))$. For
each $k\in {B}$ the equation (\ref{5.7}) is equivalent to the
boundary value problem (\ref{5.1}), (\ref{5.2}). Namely, for
each solution of the equation (\ref{5.7}) the function
$u=T_{12}(k)g$ is a solution to the boundary value problem
(\ref{5.1}), (\ref{5.2}), and for each solution $u$ to the
boundary value problem (\ref{5.1}), (\ref{5.2}) there exists the
unique solution $g$ of the equation (\ref{5.7}) related with $u$
by the equality $u=T_{12}(k)g$. The belonging
$(I+T_{13}(\cdot))^{-1}\in \mathcal{M}(L_2(\Pi_a^*),
L_2(\Pi_a^*))$ holds true.
\end{lemma}

As in the third section, we are interesting in the behaviour of
the operator $(I+T_{13}(k))^{-1}$ for small $k$, namely, we are
interested in the presence of the pole at the point $k=0$. As
the next statement shows, in distinction to  Lemma~\ref{lm3.5},
here the answer is always negative.

\begin{lemma}\label{lm5.3}
If the vicinity ${B}$ of the zero is small enough, then
$(I+T_{13}(\cdot))^{-1}\in \mathcal{H}(L_2(\Pi_a^*),
L_2(\Pi_a^*))$.
\end{lemma}
\begin{proof}
Clear, it is sufficient to prove the absence of the pole of the
operator $(I+T_{13}(k))^{-1}$. The presence of pole is
equivalent to the existence of a nontrivial solution of the
equation (\ref{5.7}) as $k=0$, $f=0$. The latter is equivalent
to the presence of nontrivial solution to the boundary value
problem (\ref{5.1}) meeting the asymptotics formulas
(\ref{5.2}). Suppose that there exists such solution to the
boundary value problem  (\ref{5.1}) and denote it by $U(x)$. The
function $U$ can be chosen being real-valued. Moreover, at the
point $x=0$  the function $U$ possess the following asymptotic
behaviour
\begin{equation*}
U(x)=\alpha r^{1/2}\sin\frac{\theta}{2}+\mathcal{O}(r),\quad
\frac{\partial}{\partial x_i}U(x)=\alpha
\frac{\partial}{\partial x_i}
r^{1/2}\sin\frac{\theta}{2}+\mathcal{O}(r),\quad r\to0,
\end{equation*}
where $(r,\theta)$ are polar coordinates associated with $x$.
These asymptotics representations can be proven in analogy with
Lemma~\ref{lm4.2}. Integrating by parts and taking into account
these asymptotics and (\ref{5.2}), we obtain
\begin{equation*}
0=\int\limits_{\Pi^*} x_1 U(\Delta+\varkappa^2)\frac{\partial
U}{\partial x_1}\,dx= 2\int\limits_{\Pi^*}\left|\frac{\partial
U}{\partial x_1}\right|^2\,dx.
\end{equation*}
This implies that the function $U$ is independent on $x_1$, what
in view of the asymptotics representations (\ref{5.2}) taken as
$x_1\to+\infty$ leads us to the equality $U=0$. The proof is
complete.
\end{proof}

As it was mentioned in the beginning of the section, the
eigenvalues of the operator $H_l$ coincides with those of a pair
of boundary value problems
\begin{equation}\label{5.8}
-\Delta \Psi=\lambda\Psi,\quad x\in\Pi^{*,l},\qquad \Psi=0,\quad
x\in\partial\Pi^{*,l}\setminus K_l,\qquad pu=0,\quad x\in K_l,
\end{equation}
where $K_l:=\{x: x_1=-l, x_2\in(-d,0)\}$, $p$ is a boundary
operator which is $pu=u$ or $pu=\frac{\partial u}{\partial
x_1}$. As $x_1\to\pm\infty$ a function $\Psi$ is assumed to meet
the asymptotics representations (\ref{5.2}) with
$k=\sqrt{\lambda-\varkappa^2}$. The eigenfunctions of the
operator $H_l$ are related with those of the boundary value
problem (\ref{5.8}) by the equalities
$\Psi_m(x)=\psi_m(x_1+l,x_2)$, $x_1>-l$, this is why the
boundary operator $p$ gives the Dirichlet condition in the case
of odd on $x_1$ functions $\psi_m(x)$ and Neumann condition in
the case of even on $x_1$ functions $\psi_m(x)$.

Our main aim at this stage is to reduce the boundary value
problem (\ref{5.8}) to an operator equation similar to
(\ref{5.7}). In order to do it we again employ the approach
which allowed us to get the equation (\ref{5.7}). We start with
the case $pu=u$. Suppose that $l>a$. We define the function
$v_0^l$ as a solution to the boundary value problem
\begin{gather*}
 -\Delta v_0^l=(\varkappa^2+k^2)v_0^l,\quad x\in\Omega_0^l,
 \\
v_0^l=0,\quad x\in\partial\Omega_0^l\setminus K_l,\qquad
v_0^l=-v_0,\quad x\in K_l.
\end{gather*}
A solution of such problem in view of the formula (\ref{5.4}) is
of the form
\begin{align}
&v_0^l(x,k)=\sum\limits_{j=1}^\infty \frac{\mathrm{e}^{-s_j^0
l}}{\sh s_j^0 l}  \beta_j(k)[g]\sh s_j^0 x_1 \sin \varkappa
j(x_2-\pi),\label{5.10}
\\
&\beta_j(k)[g]=-\frac{2}{(\pi+d) s_j^0} \int\limits_{\Omega_0^a}
g(x)\sh s_j^0 x_1 \sin j\varkappa(x_2-\pi)\,dx.\nonumber
\end{align}
The first term of this series contains the function $\sh
s_j^0l=\mathrm{i}\sin kl$ in the denominator. This function
vanishes as $kl=\pi q$, $q\in \mathbb{Z}$. At the same time, in
accordance with Item~\ref{th1.1.it1} of Theorem~\ref{th1.1} the
values $k$ corresponding to the eigenvalues of the operator
$H_l$ lie strictly inside the intervals $(\frac{\pi (m-1)}{2l},
\frac{\pi m}{2l})$. This is why the values $k=\frac{\pi m}{2l}$
are excluded from the consideration what allows us to avoid
indefiniteness in (\ref{5.10}).

By analogy with Lemma~\ref{lm3.6} one can prove that the series
(\ref{5.10}) converge in the norm of $W_2^2(\Omega_0^l)$. We set
$v^l(x,k):=v^l_0(x,k)$, $-l<x_1<0$, $v^l(x,k):=0$, $x_1>0$. We
define the function $w^l$ as the solution to the boundary value
problem (\ref{5.5}) with the function $v^l$ in the right-hand
side, i.e., $w^l=T_{11}v^l$. A solution to the boundary value
problem (\ref{5.8}) is sought as
\begin{equation}\label{5.13}
\Psi(x):=\left(w(x)+w^l(x)\right)\chi_5(x_1)+\left(1-\chi_5(x_1)\right)
\left(v(x)+v^l(x)\right),
\end{equation}
where $v(x)$, $w(x)$ are from (\ref{5.3}), (\ref{5.5}). The
function $\Psi(x)$ satisfies the boundary conditions
(\ref{5.8}), meets the asymptotics formulas (\ref{5.2}) as
$x_1\to+\infty$ and is a solution of the equation in
(\ref{5.8}), if the operator equation
\begin{equation}\label{5.11}
g+T_{13}(k)g+T_{14}(k,l)g=0,
\end{equation}
holds true, where the operator $T_{14}(k,l)$ is defined by a
rule:
\begin{equation*}
T_{14}(k,l)g:=(v^l-w^l)\left(\Delta+\varkappa^2+k^2\right)\chi_5(x_1)+
2\left(\nabla \chi_5, \nabla (v^l-w^l)\right)_{\mathbb{R}^2}.
\end{equation*}
The equation (\ref{5.11}) is equivalent to the boundary value
problem (\ref{5.8}), what can be proved by analogy with
Lemma~\ref{lm3.6}.

Let $k=k(l)$ correspond to an eigenvalue $\lambda_m(l)$ of the
operator $H_l$ by the rule $\lambda_m(l)=\varkappa^2+k^2(l)$,
and a corresponding solution $g$ of the equation (\ref{5.11})
generates an eigenfunction $\Psi_m$ in accordance with
(\ref{5.13}). It follows from Lemma~\ref{lm1.1} that each
eigenvalue of the operator $H_l$ tends to $\varkappa$ as
$l\to+\infty$, i.e., $k(l)\xrightarrow[l\to+\infty]{}0$.
Therefore, choosing $l$ great enough, we can always make $k(l)$
to belong ${B}$ for $l$ great enough. In what follows the value
$l$ is assumed to chosen in such a way.

We denote $\widehat v(x,k):=\sin kx_1\sin \varkappa(x_2-\pi)$,
$x_1<0$, $\widehat v(x,k):=0$, $x_1>0$, $\widehat w:=T_{11}
\widehat v$,
\begin{equation*}
\widehat F:=(\widehat v-\widehat
w)\left(\Delta+\varkappa^2+k^2\right)\chi_5+ 2\left(\nabla
\chi_5, \nabla (\widehat v-\widehat w)\right)_{\mathbb{R}^2}.
\end{equation*}

\begin{lemma}\label{lm5.4}
As $k$ small enough the operator $T_{14}$ can be represented as:
\begin{equation*}
T_{14}(k,l)=T_{15}(k,l)+T_{16}(k,l),\qquad
T_{15}(k,l)g=\frac{\mathrm{e}^{-\mathrm{i}kl}}{\sin
kl}\beta_0(k)[g]\widehat F,
\end{equation*}
where the operator $T_{16}(k,l)\in
\mathcal{L}(L_2(\Pi_a^*),L_2(\Pi_a^*))$ obeys an estimate:
\begin{equation*}
\|T_{16}\|\leqslant C\mathrm{e}^{-(2\sqrt{3}\varkappa-\delta)l}.
\end{equation*}
Here $C$, $\delta$ are some constants independent on $l$,
$0<\delta<2\sqrt{3}\varkappa$.

\end{lemma}

The statement of this lemma follows easily from the definition
of the function $v_0^l$ (see (\ref{5.10})) and the definition of
the operator $T_{14}$.

In view of Lemma~\ref{lm5.4} the equation (\ref{5.11}) can be
rewritten as:
\begin{equation}\label{5.30}
g+T_{13}(k)g+T_{15}(k,l)g+T_{16}(k,l)g=0.
\end{equation}
The operator $(I+T_{13}(k))$ has the bounded inverse due to
Lemma~\ref{lm5.3}, and the operator $T_{16}(k,l)$ is
exponentially small as $l\to+\infty$ in view of
Lemma~\ref{lm5.4}. Therefore, the operator
$(I+T_{13}(k)+T_{16}(k,l))^{-1}$ has also the bounded inverse
for $l$ large enough. We apply this operator to (\ref{5.30}),
what results in:
\begin{equation}
g+\frac{\mathrm{e}^{-\mathrm{i}kl}}{\sin
kl}\beta_0(k)[g](I+T_{13}(k)+T_{16}(k,l))^{-1}\widehat
F=0.\label{5.14}
\end{equation}
It is clear that $\beta_0(k)[g]\not=0$, since otherwise it would
follow from the equation obtained that $g=0$, while $g$
corresponds to the eigenfunction $\Psi_m$. Applying now the
functional $\beta_0(k)[g]$ to (\ref{5.14}), we arrive at the
equation
\begin{equation}\label{5.15}
1+\frac{\mathrm{e}^{-\mathrm{i}kl}}{\sin
kl}\beta_0(k)\left[(I+T_{13}(k)+T_{16}(k,l))^{-1} \widehat
F\right]=0.
\end{equation}
Directly from the definition of the function $\widehat F$ and
Lemmas~\ref{lm5.3},~\ref{lm5.4} it follows the equality
\begin{equation}
\beta_0(k)\left[(I+T_{13}(k)+T_{16}(k,l))^{-1} \widehat
F\right]=
ck+O\left(k^2+\mathrm{e}^{-(2\sqrt{3}\varkappa-\delta)l}\right),\label{5.17}
\end{equation}
where $c$ is a some constant, $\delta$ is the same as in
Lemma~\ref{lm5.4}. Since $k=k(l)$ corresponds to the eigenvalue
$\lambda_m$, from Lemma~\ref{lm1.1} it follows that
$k(l)=\mathcal{O}(l^{-1})$ as $l\to+\infty$. Taking into account
this equality and the realness of $k$, we substitute
(\ref{5.17}) into (\ref{5.15}):
\begin{equation}\label{5.26}
\sin kl=\mathcal{O}(l^{-1}),
\end{equation}
what implies
\begin{equation}\label{5.27}
kl=\pi q+\mathcal{O}(l^{-1}),\quad q\in \mathbb{Z}.
\end{equation}
In view of Item~\ref{th1.1.it1} of Theorem~\ref{th1.1} the index
$m$ of the eigenvalue $\lambda_m$ must be even and a two-sided
estimate
\begin{equation*}
\frac{\pi(m-1)}{2}\leqslant\pi q\leqslant\frac{\pi m}{2}
\end{equation*}
should take place. The index $m$ being even, it follows that
$q=m/2$, what by (\ref{5.27}) and the equality
$\lambda_m(l)=\varkappa^2+k^2(l)$ proves the asymptotics
expansions (\ref{1.7}).

The case of even on $x_1$ function $\psi_m$ can be proved in the
same way. The function $v_0^l$ should be chosen as a solution to
the boundary value problem
\begin{gather*}
-\Delta v_0^l=(\varkappa^2+k^2)v_0^l,\quad x\in\Omega_0^l,
\\
v_0^l=0,\quad x\in\partial\Omega_0^l\setminus K_l,\qquad
\frac{\partial v_0^l}{\partial x_1}=-\frac{\partial
v_0}{\partial x_1},\quad x\in\partial\Omega_0^l,
\end{gather*}
which is solved by separation of variables
\begin{equation*}
v_0^l(x,k)=-\sum\limits_{j=1}^\infty \frac{\mathrm{e}^{-s_j^0
l}}{\ch s_j^0 l}  \beta_j(k)[g]\sh s_j^0 x_1 \sin \varkappa
j(x_2-\pi),
\end{equation*}
where $\beta_j(k)$ are same as in (\ref{5.10}). The other
arguments are valid till Lemma~\ref{lm5.4}, if by $v_l^0$ we
mean the function just defined. The statement of
Lemma~\ref{lm5.4} is valid as well, if by $T_{15}$ we mean the
operator
\begin{equation*}
T_{15}(k,l)g=-\frac{\mathrm{e}^{-\mathrm{i}kl}}{\cos
kl}\beta_0(k)[g]\widehat F.
\end{equation*}
The deduction of the analogue of equation (\ref{5.15}) needs no
changes. In this case it is of the form:
\begin{equation*}
1-\frac{\mathrm{e}^{-\mathrm{i}kl}}{\cos
kl}\beta_0(k)\left[(I+T_{13}(k)+T_{16}(k,l))^{-1} \widehat
F\right]=0.
\end{equation*}
Using this equation, one can easily obtain an analogue of the
equation (\ref{5.26}):
\begin{equation*}
\cos kl=\mathcal{O}(l^{-1}),
\end{equation*}
what gives the equality
\begin{equation*}
kl=\frac{\pi}{2}+\pi q+\mathcal{O}(l^{-1}),\quad q\in\mathbb{Z}.
\end{equation*}
Again due to Item~\ref{th1.1.it1} of Theorem~\ref{th1.1} the
index $m$ of the eigenvalue $\lambda_m$ should be odd and
inequalities
\begin{equation*}
\frac{\pi (m-1)}{2}\leqslant\frac{\pi}{2}+\pi q\leqslant
\frac{\pi m}{2}
\end{equation*}
should take place. This implies that $q=(m-1)/2$. It proves the
asymptotics expansions in the case of even on $x_1$
eigenfunction $\psi_m$. The proof of Item~\ref{th1.2.it1} of
Theorem~\ref{th1.2} is complete.

We proceed to the proof of Item~\ref{th1.2.it2} of
Theorem~\ref{th1.2}. Let $\xi$ be an arbitrary point of the
segment $[\varkappa,1)$. For each value $l$ we choose the number
$m=m(l,\xi)$ so that the belonging
$\xi\in[\Lambda_{m-1},\Lambda_m)$ be valid. Then by
Item~\ref{th1.1.it1} of Theorem~\ref{th1.1} the estimate
\begin{equation*}
|\lambda_{m(l,\xi)}(l)-\xi|\leqslant
\Lambda_m-\Lambda_{m-1}=\frac{\pi^2(2m-1)}{4l^2}
\end{equation*}
takes place. Since $\Lambda_{m-1}\leqslant\xi<1$, it follows
that
\begin{equation*}
m\leqslant 1+\frac{2l\sqrt{1-\varkappa^2}}{\pi}.
\end{equation*}
Two last estimates yield:
\begin{equation*}
|\lambda_{m(l,\xi)}(l)-\xi|\leqslant
\frac{\pi^2}{4l^2}\left(1+\frac{4l\sqrt{1-\varkappa^2}}{\pi}\right),
\end{equation*}
what implies that $\lambda_{m(l,\xi)}\to\xi$ as $l\to+\infty$.
The proof of Theorem~\ref{th1.2} is complete.

The author thanks R.~Gadyl'shin, P.~Exner and T.~Weidl for
discussion of the work and useful remarks.

\renewcommand{\refname}


{\large References}
\begin{thebibliography}{99}

\bibitem{ESTV}
P.~Exner, P.~\v Seba, M.~Tater, D.~Van\v ek. Bound states and
scattering in quantum waveguides coupled laterally through a
boundary window /\!/ J. Math. Phys. 1996. V. 37. No. 10. P.
4867-4887.

\bibitem{HTW} Y.~Hirayama, Y.~Tokura, A.D.~Wieck, S.~Koch,
R.J.~Haug, K.~von~Klitzing, K.~Ploog. Transport characteristics
of a window-coupled in-plane-gated wire system /\!/ Physical
Review B. 1993. V. 48. No. 11. P. 7991-7998.

\bibitem{BGRS}
W.~Bulla, F.~Gesztesy, W.~Renger, B.~Simon. Weakly coupled bound
states in quantum waveguides /\!/ Proc. Amer. Math. Soc. 1997.
V. 125. No. 5. P. 1487-1495.

\bibitem{EV1} P.~Exner and S.~Vugalter. Asymptotics estimates
for bound states in quantum waveguides coupled laterally through
a narrow window /\!/ Ann. Inst. H. Poincare. 1996. V. 65.  No.
1. P. 109-123.

\bibitem{EV2} P.~Exner and S.~Vugalter. Bound-state asymptotic
estimate for window-coupled Dirichlet strips and layers /\!/ J.
Phys. A. 1997. V. 30. No. 22. P. 7863-7878.

\bibitem{P} I.Yu.~Popov. Asymptotics of bound states for
laterally coupled waveguides /\!/ Reports on Mathematical
Physics. 1999. V. 43. No. 3. P. 427-437.

\bibitem{G} R. Gadyl'shin. On regular and singular perturbation
of acoustic and quantum waveguides /\!/ Comptes Rendus
Mechanique. 2004. V. 332. No. 8. P. 647-652.

\bibitem{BEG}
D. Borisov, P. Exner and R. Gadyl'shin. Geometric coupling
thresholds in a two-dimensional strip /\!/ J. Math. Phys. 2002.
V. 43. No. 12. P. 6265-6278.

\bibitem{Ku} Ch. Kunze. Leaky and mutually coupled wires /\!/
Physical Review B. 1993. V. 48. No. 19. P. 14338-14346.

\bibitem{DK} J.~Dittrich and J.~K\v{r}\'\i\v{z}. Bound states in
straight quantum waveguide with combined boundary condition /\!/
J. Math. Phys. 2002. V. 43. No. 8. P. 3892-3915.

\bibitem{BEK} D.~Borisov, T.~Ekholm and H.~Kova\v r\'\i k.
Spectrum of the magnetic Schr\"odinger operator in a waveguide
with combined boundary conditions /\!/ Ann. H. Poincare. 2005.
V. 6. No. 2. P. 327-342.

\bibitem{B}
M.S. Birman. Perturbation of the continuous spectrum of a
singular elliptic operator under a change of the boundary and
the boundary condition /\!/ Vestnik Leningradskogo universiteta.
1962. V. 17. No. 1. P. 22-55.



\bibitem{Ad} R.A.~Adams. Sobolev Spaces. N.Y.: Academic Press.
1975.

\bibitem{Ld} V.P.~Mikhajlov. Partial differential equations.
Moscow: Mir Publishers, 1978.

\bibitem{BEH}
J.~Blank, P.~Exner and M.~Havli\v{c}ek. Hilbert Space Operators
in Quantum Physics. N.Y.: AIP Press, 1994.

\bibitem{K} T.~Kato.
Perturbation theory for linear operators. N.Y.: Springer-Verlag,
1966.


\bibitem{RS}
M.~Reed, B.~Simon. Methods of modern mathematical physics. I\!V:
Analysis of operators. N.Y.: Academic Press, 1978.

\bibitem{BE}
D.~Borisov and P.~Exner. Exponential splitting of bound states
in a waveguide with a pair of distant windows /\!/ J. Phys. A.
2004. V. 37. No. 10. P. 3411-3428.

\bibitem{SP} E.~Sanchez-Palencia. Homogenization Techniques for
Composite Media. Berlin-New York: Springer-Verlag, 1987.

\bibitem{G1} Gadyl'shin~R.R.
Local Perturbations of the Schr\"odinger Operator on the Axis
/\!/ Theor. Math. Phys. 2002. V. 132. No. 1. P. 976-982.

\bibitem{G2}
R.R. Gadyl'shin. On local perturbations of the Schroedinger
operator on the plane /\!/ Theor. Math. Phys. 2004. V. 138. No.
1. P. 33-44.

\bibitem{Il} A.M.~Il'in. Matching of Asymptotic Expansions
of Solutions of Boundary Value Problems. Amer. Mat. Soc.,
Providence, RI, 1992.

\end{thebibliography}
\end{document}